\documentclass[12pt]{article}
\usepackage{epsfig}

\newcommand{\mysection}{\setcounter{equation}{0}\section}

\def\beq{\begin{equation}}
\def\eeq{\end{equation}}
\def\beqa{\begin{eqnarray}}
\def\eeqa{\end{eqnarray}}
 
\newlength{\dinwidth} \newlength{\dinmargin}
\begin{document}

\begin{center}
{\Large \bf Next-to-next-to-leading soft-gluon corrections for the top quark cross section and transverse momentum distribution}
\end{center}
\vspace{2mm}
\begin{center}
{\large Nikolaos Kidonakis}\\
\vspace{2mm}
{\it Kennesaw State University,  Physics \#1202,\\
1000 Chastain Rd., Kennesaw, GA 30144-5591, USA}
\end{center}
 
\begin{abstract}
I present results for top quark production in hadronic collisions 
at LHC and Tevatron energies. The soft-gluon corrections to the 
differential cross section are resummed 
at next-to-next-to-leading-logarithm (NNLL) accuracy via the 
two-loop soft anomalous dimension matrices. 
Approximate next-to-next-to-leading-order (NNLO) differential and total 
cross sections are calculated.
Detailed theoretical predictions are shown for the $t{\bar t}$ cross section 
and the top quark $p_T$ distribution at the Tevatron and the LHC.
\end{abstract}
 
\mysection{Introduction}

The top quark occupies a unique position in the list of elementary particles as the most massive 
particle discovered to date. Its high mass suggests an important role for the top quark in 
the physics of electroweak symmetry breaking. After a long period of searches, the discovery of the top quark 
via top-antitop production in proton-antiproton collisions  
($p {\bar p} \rightarrow t{\bar t}$) was announced in 1995 by the CDF and D0 collaborations 
at the Fermilab Tevatron collider \cite{CDFD0ttbar}.  
The $t{\bar t}$ cross section has been measured with increasing precision at Run II of the Tevatron \cite{CDFcs,D0cs}
and there has also been data for the transverse momentum, $p_T$, distribution of the top quark \cite{TevpT}.
More recently single top quark production was observed by D0 \cite{D0st} and CDF \cite{CDFst}.
Measurements of the top quark mass have also been increasingly more precise \cite{topmass}.
The LHC is expected to observe a very large number of top quark events and to bring top quark physics to a new energy 
frontier. 
For reviews of top quark physics at the Tevatron and the LHC see 
Ref. \cite{toprevex} (experiment) and Ref. \cite{toprevph} (theory).

The experimental measurements of the top quark cross section and $p_T$ distribution at the Tevatron 
are currently in good agreement with theoretical predictions. However, as the experimental errors 
continue to get smaller with time, precise theoretical calculations with smaller uncertainties are required. 
Next-to-leading order (NLO) calculations of the QCD corrections have been available for over two decades \cite{NLO1,NLO2} 
(electroweak corrections, which are much smaller numerically, have also been calculated more recently \cite{topew}) 
but the associated uncertainty is much bigger than current experimental errors. 
The inclusion of higher-order soft-gluon corrections enhances the cross 
section and $p_T$ distribution and significantly reduces the theoretical error \cite{NKRV1,NKRV2}. 

Until recently, the state of the art in theoretical predictions was approximate next-to-next-to-leading order (NNLO) 
calculations based on next-to-leading-logarithm (NLL) resummation of 
soft-gluon corrections for the differential cross section, supplemented with further subleading terms 
\cite{NKRV1,NKRV2}. These soft-gluon corrections are dominant not only near partonic threshold but also away from it. 
The accuracy at NLL was achieved by the calculation of the one-loop soft anomalous dimension matrices for 
the partonic channels in top quark production in Ref. \cite{NKGS}. 

To achieve next-to-next-to-leading-logarithm (NNLL) accuracy in the resummation one needs to calculate the 
soft anomalous dimensions at two loops. This is a much more difficult undertaking.
For massless quark scattering, the two-loop soft anomalous dimension matrix was first calculated in Ref. \cite{ADS}.
Further work on soft and collinear singularities of dimensionally-regularized scattering amplitudes in 
massless gauge theories followed in Refs. \cite{DMS,BN,GM,LD,DGM}.  
More recently, a lot of work on massive two-loop soft anomalous dimensions has 
appeared in Refs. \cite{NK2l,MSS,BNm,BFS,CMS,FNPY,NK2ls,NK2lw,AFNPY,MSS2,BFS2}. The presence of a mass for the top quark considerably 
complicates the calculation relative to the massless case.

Soft-gluon resummation is a consequence of factorization. The partonic cross 
section can be factorized into functions associated with the hard scattering, 
collinear and soft-gluon emission from the external partons, and noncollinear 
soft-gluon emission that depends on the color structure of the process \cite{NKGS}. 
The renormalization group evolution of these functions results in expressions 
for the resummed cross section. The resummation formalism followed here has already been 
presented and reviewed in numerous papers over a period of more than a decade 
(see Refs. \cite{NKRV1,NKGS,NK2ls,NK2lw,NK01,KLMV,NNNLO} and references therein) so we will not 
repeat the derivation of resummation and we will not repeat explicit expressions in this paper except for 
new two-loop results in Section 2. Resummation is performed 
in Mellin moment space: we define a 
kinematical variable $s_4$ that measures distance from partonic threshold, and then $N$ 
is the moment variable conjugate to $s_4$. 
For $t{\bar t}$ production the resummed partonic cross section
in moment space is given by 
\beqa
&& {\hat{\sigma}}^{res}(N) =   
\exp\left[\sum_{i=a,b} E_i(N_i)\right] \;  
\exp \left[ 2 \sum_{i=a,b} \int_{\mu_F}^{\sqrt{s}} \frac{d\mu}{\mu}\;
\gamma_{i/i}\left({\tilde N}_i, \alpha_s(\mu)\right)\right] \;
\rm{Tr}\left\{H_{ab}\left(\alpha_s(\sqrt{s})\right) \right.
\nonumber\\ && \hspace{-5mm} \left. \times \, 
\exp \left[\int_{\sqrt{s}}^{{\sqrt{s}}/{\tilde N'}} 
\frac{d\mu}{\mu}\; \Gamma_{S\, ab}^{\dagger}\left(\alpha_s(\mu)\right)\right]\;
S_{ab}\left(\alpha_s(\sqrt{s}/{\tilde N'})\right) \; 
\exp \left[\int_{\sqrt{s}}^{{\sqrt{s}}/{\tilde N'}} 
\frac{d\mu}{\mu}\; \Gamma_{S\, ab}\left(\alpha_s(\mu)\right)\right]\right\} 
\,. 
\label{resHS}
\eeqa
The first exponent in the above expression 
resums soft and collinear corrections from the incoming partons $a$ and $b$ 
(quark-antiquark or gluon-gluon)
while the second exponent controls the factorization scale, $\mu_F$, 
dependence of the cross section. 
$H_{ab}$ is the hard-scattering function while $S_{ab}$ is the 
soft function describing noncollinear soft gluon emission.
The renormalization group evolution of the soft function is controlled by 
the soft anomalous dimension, $\Gamma_{S\, ab}$ \cite{NKGS}.
It is important to note that $H_{ab}$, $S_{ab}$, and $\Gamma_{S\, ab}$ are 
matrices in the space of color structures of the process \cite{NKGS,NK01,KLMV}.
In the next section we will present 
explicit expressions for the new 
two-loop results for the soft anomalous dimension matrices 
$\Gamma_{S\, q{\bar q}}$, for the 
$q{\bar q} \rightarrow t{\bar t}$ channel, and $\Gamma_{S\, gg}$, for the 
$gg \rightarrow t{\bar t}$ channel. It is these new ingredients that allow us 
to complete the NNLL resummation in our formalism 
(for other approaches see Refs. \cite{BFS,CMS,FNPY,AFNPY,BFS2} and the discussion in Section 5).

The resummed cross section, Eq. (\ref{resHS}) can be expanded at fixed order in $\alpha_s$ 
to NLO, NNLO, etc.,  and inverted back to momentum space, see e.g. Refs. \cite{NKRV1,NK01,KLMV,NNNLO}. 
At each order in $\alpha_s$, one encounters plus-distribution terms of the form 
$[\ln^k(s_4/m^2)/s_4]_+$, where $m$ is the top quark mass and, for the $n$-th order corrections, 
the power of the logarithm, $k$, can range from the leading value of $2n-1$  down to the lowest value of 0. 
Thus at NLO $k=1$ or $0$, while at NNLO $k$ can take the values 
3,2,1,0. From NLL resummation one can determine the coefficients of both, $k=1,0$, powers of the 
logarithms at NLO, but only the powers $k=3,2,1$ at NNLO (determining the NNLO $k=1$ term requires matching with NLO).
Partial results for the $k=0$ term at NNLO were provided in Ref. \cite{NKRV1} and also used in \cite{NKRV2}.
From NNLL resummation one can in addition fully determine the $k=0$ term at NNLO.

In the following section we present the soft anomalous dimension matrices for 
the $q{\bar q} \rightarrow t{\bar t}$ and $gg \rightarrow t{\bar t}$ channels 
at one and two loops. In Section 3 we use the NNLL resummation to obtain approximate 
NNLO results for the total $t{\bar t}$ cross section and the top quark 
$p_T$ distribution in proton-antiproton collisions at the Tevatron. In Section 
4 corresponding results are given for proton-proton collisions at LHC energies. 
A comparison with other approaches and conclusions are given in Section 5.

\mysection{Soft anomalous dimension matrices for $t {\bar t}$ production}

We begin with the result for the soft (cusp) anomalous dimension 
$\Gamma_S$ \cite{NK2l} for $e^+ e^- \rightarrow t {\bar t}$, 
which is an integral part of the calculation for the soft anomalous 
dimension matrices $\Gamma_{S\, q{\bar q}}$  and $\Gamma_{S\, gg}$ 
for $t {\bar t}$ hadroproduction.  
The calculations of soft anomalous dimensions involve diagrams 
with eikonal lines representing the top quarks.  
The eikonal diagrams are calculated in Feynman gauge in momentum space, 
and we use dimensional regularization with 
$d=4-\epsilon$ dimensions to isolate the ultraviolet (UV) poles of the diagrams.
The soft anomalous dimension is then determined from 
the coefficients of the UV poles \cite{NK2l}. 
Writing 
$\Gamma_S=(\alpha_s/\pi) \Gamma_S^{(1)}+(\alpha_s/\pi)^2 \Gamma_S^{(2)}
+\cdots$, we have the one-loop expression 
\beq
\Gamma_S^{(1)}=C_F \left[-\frac{(1+\beta^2)}{2\beta} 
\ln\left(\frac{1-\beta}{1+\beta}\right) -1\right]=-C_F \left[L_{\beta}+1\right]
\label{GammaS1}
\eeq
where $C_F=(N_c^2-1)/(2N_c)$, with $N_c=3$ the number of colors; $\beta=\sqrt{1-4m^2/s}$,  
with $s$ the squared c.m. energy; and 
\beq
L_{\beta}=\frac{1+\beta^2}{2\beta}\ln\left(\frac{1-\beta}{1+\beta}\right)\, .
\eeq

The two-loop soft (cusp) anomalous dimension, determined from the UV poles of two-loop eikonal diagrams, 
is \cite{NK2l,NK2lw}
\beq
\Gamma_S^{(2)}=\frac{K}{2} \, \Gamma_S^{(1)}
+C_F C_A M_{\beta}
\label{Gammas2}
\eeq
where 
$K=C_A (67/18-\zeta_2)-5n_f/9$, with $C_A=N_c$ and $n_f=5$ the number of light-quark flavors.
We have written the two-loop result $\Gamma_S^{(2)}$ 
in Eq. (\ref{Gammas2}) in the form of a term which is a multiple 
of the one-loop soft anomalous dimension $\Gamma_S^{(1)}$ 
plus additional terms, denoted as $M_{\beta}$:
\beqa
&& M_{\beta}=\frac{1}{2}+\frac{\zeta_2}{2}
+\frac{1}{2} \ln^2\left(\frac{1-\beta}{1+\beta}\right)
\nonumber \\ && \hspace{-10mm}
{}-\frac{(1+\beta^2)^2}{8 \beta^2} \left[\zeta_3
+\zeta_2 \ln\left(\frac{1-\beta}{1+\beta}\right)
+\frac{1}{3} \ln^3\left(\frac{1-\beta}{1+\beta}\right)
+\ln\left(\frac{1-\beta}{1+\beta}\right) 
{\rm Li}_2\left(\frac{(1-\beta)^2}{(1+\beta)^2}\right) 
-{\rm Li}_3\left(\frac{(1-\beta)^2}{(1+\beta)^2}\right)\right] 
\nonumber \\ &&  \hspace{-10mm}
{}-\frac{(1+\beta^2)}{4 \beta} \left[\zeta_2
-\zeta_2 \ln\left(\frac{1-\beta}{1+\beta}\right) 
+\ln^2\left(\frac{1-\beta}{1+\beta}\right)
-\frac{1}{3} \ln^3\left(\frac{1-\beta}{1+\beta}\right) \right.
\nonumber \\ &&  \hspace{45mm} \left. 
{}+2  \ln\left(\frac{1-\beta}{1+\beta}\right) 
\ln\left(\frac{(1+\beta)^2}{4 \beta}\right) 
-{\rm Li}_2\left(\frac{(1-\beta)^2}{(1+\beta)^2}\right)\right] \, .
\nonumber \\
\label{Mbeta}
\eeqa
This result, first obtained in \cite{NK2l}, is written in terms of logarithms, 
dilogarithms, and trilogarithms, and it provides a more explicit analytical 
expression than earlier work \cite{KorRad}.
Note that as $\beta \rightarrow 1$, $M_{\beta} \rightarrow (1-\zeta_3)/2$.

We can now proceed with the results for the two-loop soft anomalous 
dimension matrices for the partonic processes 
$q {\bar q} \rightarrow t{\bar t}$ and $gg \rightarrow t{\bar t}$. 
The calculation involves the two-loop soft (cusp) anomalous dimension for all 
pairs of external lines in the process (cf. \cite{NK2l,NK2ls,NK2lw}) as well as graphs with 
gluons connecting three external lines (cf. \cite{MSS,FNPY,MSS2}).
We begin with top quark
production through light quark annihilation,
\beq
q(p_a)+{\bar q}(p_b) \rightarrow t(p_1) + {\bar t}(p_2)\, .
\eeq
We define the kinematical invariants
\beq
s=(p_a+p_b)^2\, , \quad t_1=(p_b-p_1)^2-m^2\, , \quad u_1=(p_a-p_1)^2-m^2\, ,
\label{stu}
\eeq
and $s_4=s+t_1+u_1$, where $s_4$ measures distance from partonic threshold.
The calculations are performed in a color tensor basis
consisting of singlet and octet exchange in the $s$ channel,
\beq
c_1 = \delta_{ab}\delta_{12}\, , \quad \quad
c_2 =  T^c_{F\; ba} \, T^c_{F\; 12}\, .
\label{c1c2}
\eeq
Here the color indices for the incoming (light) quark and antiquark are $a$ and $b$,
respectively, and for the outgoing top quark and antiquark $1$ and $2$, 
respectively, and $T^c_F$ are the generators of $SU(3)$ in the fundamental representation.   

The matrix for $q{\bar q} \rightarrow t{\bar t}$ in this $c_1$, $c_2$ color basis is 
\beq
\Gamma_{S\, q{\bar q}}=\left[\begin{array}{cc}
\Gamma_{q{\bar q} \, 11} & \Gamma_{q{\bar q} \, 12} \\
\Gamma_{q{\bar q} \, 21} & \Gamma_{q{\bar q} \, 22}
\end{array}
\right] \, .
\label{matrixqqtt}
\eeq

At one loop:
\beqa
\Gamma_{q{\bar q} \,11}^{(1)}&=&-C_F \, [L_{\beta}+1]=\Gamma_S^{(1)} \, ,
\nonumber \\
\Gamma_{q{\bar q} \,21}^{(1)}&=&
2\ln\left(\frac{u_1}{t_1}\right) \, ,
\nonumber \\
\Gamma_{q{\bar q} \,12}^{(1)}&=&
\frac{C_F}{C_A} \ln\left(\frac{u_1}{t_1}\right) \, ,
\nonumber \\
\Gamma_{q{\bar q} \,22}^{(1)}&=&C_F
\left[4\ln\left(\frac{u_1}{t_1}\right)
-L_{\beta}-1\right]
+\frac{C_A}{2}\left[-3\ln\left(\frac{u_1}{t_1}\right)
+\ln\left(\frac{t_1u_1}{s m^2}\right)+L_{\beta}\right]\, .
\label{Gamma1qqtt}
\eeqa

The result in Eq. (\ref{Gamma1qqtt}) is somewhat different from the original in 
Ref. \cite{NKGS} because the original calculation used the axial gauge while 
Eq. (\ref{Gamma1qqtt}) is in Feynman gauge. Of course this does not affect the 
complete resummed expression because other terms in the resummed 
cross section compensate 
by also taking different forms in the two gauges.
We note that the ``11'' element of the matrix is simply the cusp 
anomalous dimension, $\Gamma_S$. 

At two loops:
\beqa
\Gamma_{q{\bar q} \,11}^{(2)}&=&\frac{K}{2} \Gamma_{q{\bar q} \,11}^{(1)}
+C_F C_A \, M_{\beta}=\Gamma_S^{(2)}\, ,
\nonumber \\
\Gamma_{q{\bar q} \,21}^{(2)}&=&
\frac{K}{2}  \Gamma_{q{\bar q} \,21}^{(1)} +C_A N_{\beta} \ln\left(\frac{u_1}{t_1}\right) \, ,
\nonumber \\
\Gamma_{q{\bar q} \,12}^{(2)}&=&
\frac{K}{2} \Gamma_{q{\bar q} \,12}^{(1)} -\frac{C_F}{2} N_{\beta} \ln\left(\frac{u_1}{t_1}\right) \, ,
\nonumber \\
\Gamma_{q{\bar q} \,22}^{(2)}&=&
\frac{K}{2} \Gamma_{q{\bar q} \,22}^{(1)}
+C_A\left(C_F-\frac{C_A}{2}\right) \, M_{\beta} \, .
\label{Gamma2qqtt}
\eeqa
Here the term 
\beqa
N_{\beta}&=&\frac{1}{2}\ln^2\left(\frac{1-\beta}{1+\beta}\right)
-\frac{(1+\beta^2)}{4 \beta} \left[
\ln^2\left(\frac{1-\beta}{1+\beta}\right)
+2  \ln\left(\frac{1-\beta}{1+\beta}\right) 
\ln\left(\frac{(1+\beta)^2}{4 \beta}\right) 
-{\rm Li}_2\left(\frac{(1-\beta)^2}{(1+\beta)^2}\right)\right]
\nonumber \\
\eeqa
comes from graphs with gluons connecting three external lines, whose contribution 
were first calculated explicitly in \cite{FNPY}.
Note that $N_{\beta}$ is just a subset of the terms of $M_{\beta}$, 
Eq. (\ref{Mbeta}), so all analytical structures already appear 
in $M_{\beta}$,  and that 
as $\beta \rightarrow 1$, $N_{\beta} \rightarrow 0$.
The two-loop matrix, Eq. (\ref{Gamma2qqtt}) is not proportional to the one-loop matrix, 
Eq. (\ref{Gamma1qqtt}).
This fact was first discussed in Ref. \cite{NK2l} and it is to be contrasted with the simple 
proportionality relation for the massless case that was found in Ref. \cite{ADS}.

We continue with the $gg$ channel:
\beq
g(p_a)+g(p_b) \rightarrow t(p_1) + {\bar t}(p_2)\, .
\eeq
We choose the following basis for the color factors:
\beq
c_1=\delta^{ab}\,\delta_{12}, \quad c_2=d^{abc}\,T^c_{12},
\quad c_3=i f^{abc}\,T^c_{12} 
\eeq
where $d^{abc}$ and $f^{abc}$ are the totally symmetric and 
antisymmetric $SU(3)$ invariant tensors, respectively.
We define $s$, $t_1$, and $u_1$ for this channel as in Eq. (\ref{stu}).

The matrix for $gg \rightarrow t{\bar t}$ in this basis is 
\beq
\Gamma_{S \, gg}=\left[\begin{array}{ccc}
\Gamma_{gg\, 11} & 0 & \Gamma_{gg\,13} \vspace{2mm} \\
0 & \Gamma_{gg\,22} & \Gamma_{gg\,23} \vspace{2mm} \\
\Gamma_{gg\,31} & \Gamma_{gg\,32} & \Gamma_{gg\,22}
\end{array}
\right] \, .
\label{matrixggtt}
\eeq

At one loop:
\beqa
\Gamma_{gg\, 11}^{(1)}&=&-C_F [L_{\beta}+1]=\Gamma_S^{(1)} \, ,
\nonumber \\
\Gamma_{gg\, 31}^{(1)}&=&2 \ln\left(\frac{u_1}{t_1}\right) \, ,
\nonumber \\
\Gamma_{gg\, 13}^{(1)}&=& \ln\left(\frac{u_1}{t_1}\right) \, ,
\nonumber \\
\Gamma_{gg\, 22}^{(1)}&=&-C_F [L_{\beta}+1] 
+\frac{C_A}{2}\left[\ln\left(\frac{t_1 u_1}{m^2 s}\right)+L_{\beta}\right] \, ,
\nonumber \\
\Gamma_{gg\, 32}^{(1)}&=&\frac{N_c^2-4}{2N_c} \ln\left(\frac{u_1}{t_1}\right) \, ,
\nonumber \\
\Gamma_{gg\, 23}^{(1)}&=&\frac{C_A}{2} \ln\left(\frac{u_1}{t_1}\right) \, .
\label{Gamma1ggtt}
\eeqa
The expression in Eq. (\ref{Gamma1ggtt}) is again somewhat different from the original in 
Ref. \cite{NKGS} because  Eq. (\ref{Gamma1ggtt}) is derived in Feynman gauge. 

At two loops:
\beqa
\Gamma_{gg\, 11}^{(2)}&=& \frac{K}{2} \Gamma_{gg \,11}^{(1)}
+C_F C_A \, M_{\beta}=\Gamma_S^{(2)} \, ,
\nonumber \\
\Gamma_{gg\, 31}^{(2)}&=&\frac{K}{2} \Gamma_{gg \,31}^{(1)} 
+C_A N_{\beta} \ln\left(\frac{u_1}{t_1}\right) \, ,
\nonumber \\
\Gamma_{gg\, 13}^{(2)}&=&\frac{K}{2} \Gamma_{gg \,13}^{(1)} 
-\frac{C_A}{2} N_{\beta} \ln\left(\frac{u_1}{t_1}\right) \, ,
\nonumber \\
\Gamma_{gg\, 22}^{(2)}&=& \frac{K}{2} \Gamma_{gg \,22}^{(1)}
+C_A \left(C_F-\frac{C_A}{2}\right) \, M_{\beta} \, ,
\nonumber \\
\Gamma_{gg\, 32}^{(2)}&=&\frac{K}{2} \Gamma_{gg \,32}^{(1)} \, , 
\nonumber \\
\Gamma_{gg\, 23}^{(2)}&=&\frac{K}{2} \Gamma_{gg \,23}^{(1)} \, .
\label{Gamma2ggtt}
\eeqa
As was the case for the $q{\bar q}$ channel, we note that for the $gg$ channel the two-loop matrix, 
Eq. (\ref{Gamma2ggtt}) is not proportional to the one-loop matrix, Eq. (\ref{Gamma1ggtt}).

The expressions in Eqs. (\ref{Gamma2qqtt}) and (\ref{Gamma2ggtt}) are different from the 
corresponding ones in \cite{FNPY} due to different definitions and formalism. 

With the two-loop soft anomalous dimension matrices at hand we 
achieve NNLL accuracy in the resummed cross section, Eq. (\ref{resHS}). 
Expanding the resummed cross section to NNLO we then calculate 
approximate NNLO cross sections and transverse momentum distributions 
for top quarks at the Tevatron and the LHC.

\mysection{Top cross section and $p_T$ distribution at the Tevatron}

We now provide a detailed phenomenological study of top quark 
production at the Tevatron collider, including the total 
$t {\bar t}$ cross section and the top quark $p_T$ distribution. 
We present NLO and approximate NNLO calculations for these quantities. 
The NNLO approximate results are computed by adding the NNLO soft-gluon 
corrections (derived from NNLL resummation) to the exact NLO quantities.
The total and differential cross sections depend on the factorization 
scale, $\mu_F$, and the renormalization scale, $\mu_R$. These two scales 
are often set equal to each other and denoted simply as $\mu$, but they are 
in principle independent.

\subsection{$t{\bar t}$ cross section at the Tevatron}

\begin{figure}
\begin{center}
\includegraphics[width=11cm]{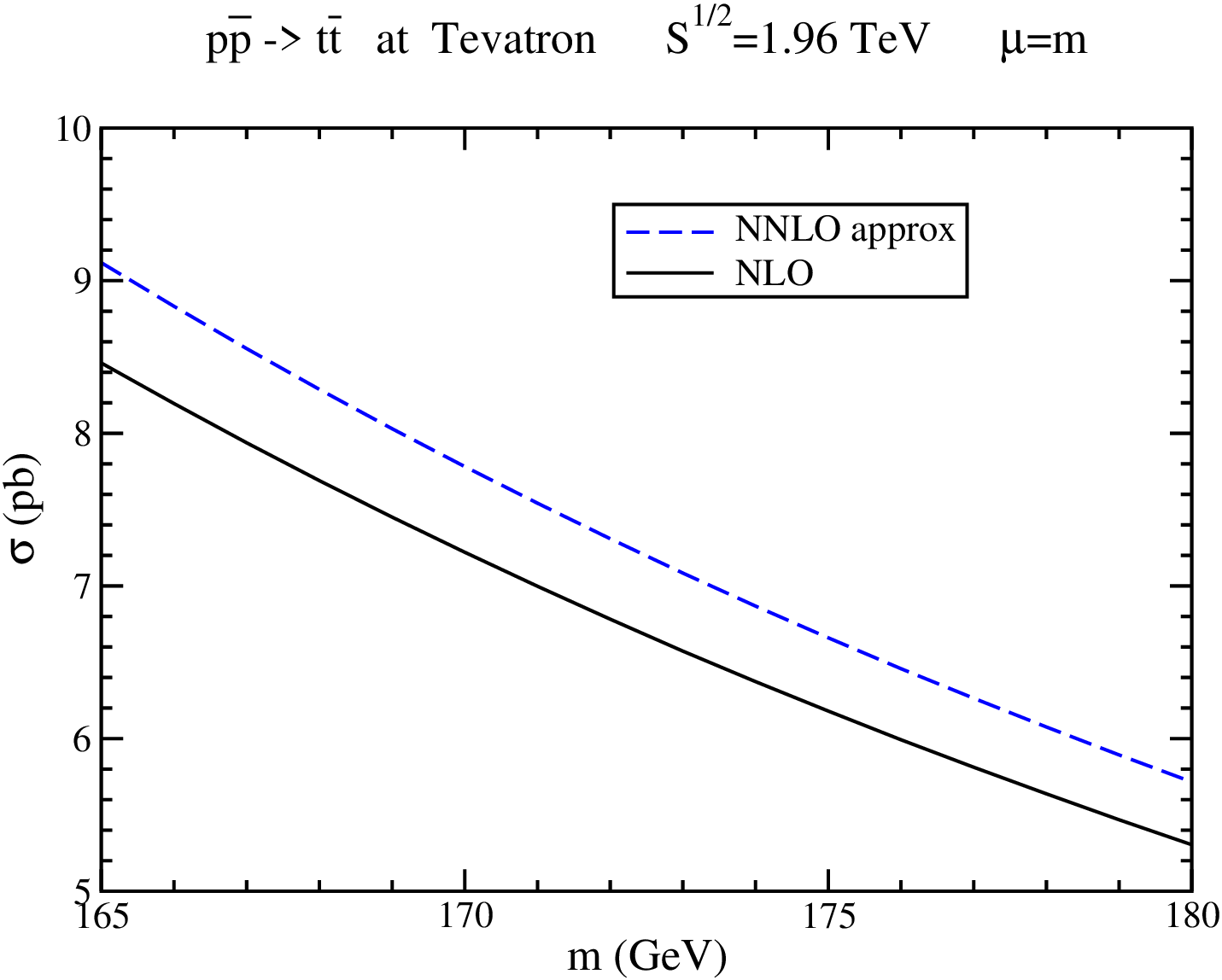}
\caption{The NLO and approximate NNLO cross section for $t{\bar t}$ production 
at the Tevatron with $\sqrt{S}=1.96$ TeV and MSTW2008 NNLO pdf.}
\label{toptev}
\end{center}
\end{figure}

In Fig. \ref{toptev} we plot the cross section for top-antitop production 
in proton-antiproton collisions at the Tevatron over a top quark mass 
range $165 \le m \le 180 $ GeV 
at a factorization and renormalization scale $\mu=m$. The exact 
NLO and the approximate NNLO cross sections are shown. The enhancement 
from the NNLO soft-gluon corrections is 7.8\%. Here  
we have used the MSTW2008 NNLO parton distibution functions (pdf) 
\cite{MSTW2008}. We will use these pdf for our calculations throughout 
this paper except where noted otherwise.

\begin{table}[htb]
\begin{center}
\begin{tabular}{|c|c|c|c|} \hline
\multicolumn{4}{|c|}{NNLO approx $t{\bar t}$ cross section (pb)} \\ \hline
$m$ (GeV) & Tevatron & LHC 7 TeV & LHC 14 TeV \\ \hline
170 & 7.78 & 179 & 998 \\ \hline 
171 & 7.54 & 173 & 972 \\ \hline 
172 & 7.31 & 168 & 946 \\ \hline 
173 & 7.08 & 163 & 920 \\ \hline 
174 & 6.87 & 158 & 896 \\ \hline 
175 & 6.66 & 154 & 873 \\ \hline 
\end{tabular}
\caption[]{The NNLO approximate $t{\bar t}$ production cross section in pb 
in $p{\bar p}$ collisions at the Tevatron with $\sqrt{S}=1.96$ TeV and   
in $pp$ collisions at the LHC with $\sqrt{S}=7$ TeV and 14 TeV. We set 
$\mu=m$ and use the MSTW2008 NNLO pdf \cite{MSTW2008}.}
\label{table1}
\end{center}
\end{table}

Table 1 lists the values for the NNLO approximate cross section at the 
Tevatron for top quark masses between 170 GeV and 175 GeV. 
There are two kinds of theoretical uncertainties associated with the 
calculation: dependence on the factorization/renormalization scale, 
and uncertainties from the parton densities. 

\begin{figure}
\begin{center}
\includegraphics[width=11cm]{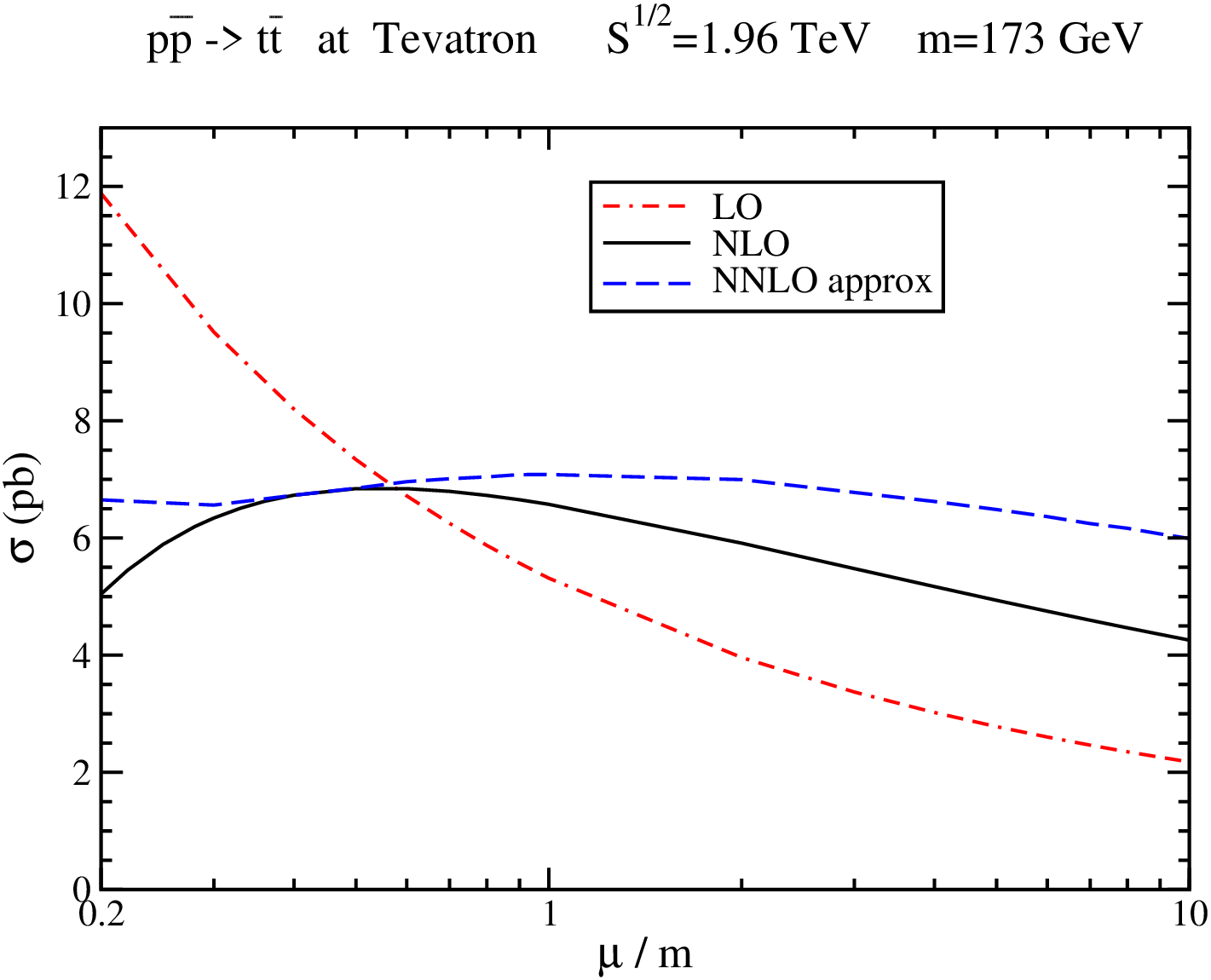}
\caption{The scale dependence of the $t{\bar t}$ cross section  
at the Tevatron with $\sqrt{S}=1.96$ TeV and $m=173$ GeV.}
\label{toptevmu}
\end{center}
\end{figure}

The scale dependence of the cross section for $m=173$ GeV 
is plotted in Fig. \ref{toptevmu}
over a range of two orders of magnitude, $0.2 \le \mu/m \le 10$.  
It is clear that at leading order (LO) the cross section is strongly dependent 
on the choice of scale, varying by a factor of 5.45 between maximum and minimum 
values in the range shown. 
The NLO corrections significantly stabilize the LO variation: the NLO cross 
section varies by a factor of 1.61. The NNLO soft-gluon corrections further 
reduce the scale dependence: the NNLO approximate cross section varies 
by a factor of only 1.18. The improvement provided by the NNLO corrections is 
even more impressive if one considers only the variation $0.5 \le \mu/m \le 2$ 
as traditionally used to estimate errors. For this range the LO 
cross section varies by a factor of 1.85, the NLO cross section by 1.16, 
while the NNLO approximate cross section by a factor of only 1.034.

For a top quark mass of 173 GeV, the NLO cross section is 
$6.57 {}^{+0.27}_{-0.66} {}^{+0.34}_{-0.25}$ pb
and the NNLO approximate cross section is
\beq
\sigma^{\rm NNLOapprox}_{t{\bar t}}(m=173 \, {\rm GeV}, \, 1.96\, {\rm TeV})
=7.08 {}^{+0.00}_{-0.24} {}^{+0.36}_{-0.27} \; {\rm pb} \, .
\eeq
Here the first uncertainty is from scale variation over $0.5 \le \mu/m \le 2$ 
and the second is from the MSTW2008 NNLO pdf errors at 90\% C.L. (to be 
conservative, we do not use the smaller 68\% C.L. pdf errors). 
At NLO the scale uncertainty is bigger than that from the pdf,  
but at NNLO the scale uncertainty is much smaller than the pdf one. 
In fact the scale uncertainty at NNLO is about four times smaller than that 
at NLO, again highlighting the dramatic reduction of scale dependence provided 
by the higher-order corrections. Adding the scale and pdf errors in quadrature,  
the NNLO approximate result is $7.08 \pm 0.36$ pb, i.e. we have a $\pm$ 5.1\% 
total uncertainty, which is to be contrasted with a much larger (+6.6\% -10.7\%) 
total error (in quadrature) at NLO. 

\begin{figure}
\begin{center}
\includegraphics[width=11cm]{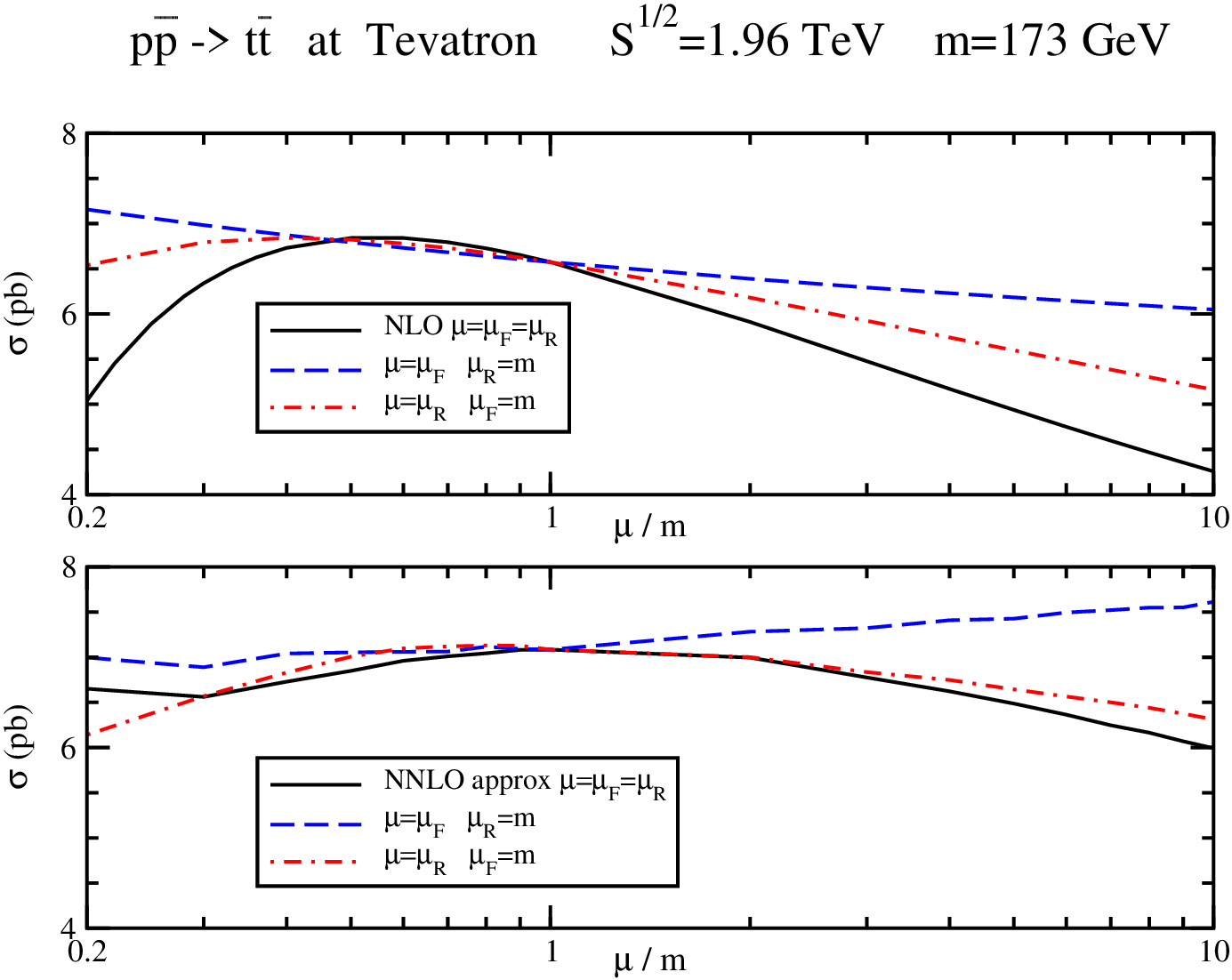}
\caption{The $\mu_F$ and $\mu_R$ dependence of the $t{\bar t}$ cross section  
at the Tevatron with $\sqrt{S}=1.96$ TeV and $m=173$ GeV. The top plot 
is at NLO and the bottom is at approximate NNLO accuracy.}
\label{toptevmufr}
\end{center}
\end{figure}

One can also study the dependence of the cross section separately on the 
factorization scale and the renormalization scale. This can be important 
because in some cases setting $\mu_F$ equal to $\mu_R$ may give a 
smaller uncertainty than from varying the scales independently.
In Fig. \ref{toptevmufr} we plot the scale dependence of the cross section 
for $m=173$ GeV in three different 
ways at NLO (top plot) and approximate NNLO (bottom plot). The first way 
is to set $\mu=\mu_F=\mu_R$ and vary this common scale, exactly as we did 
in Fig. \ref{toptevmu}. The second way is to vary the factorization scale 
$\mu_F$ while keeping the renormalization scale fixed at $\mu_R=m$. 
The third way  is to vary $\mu_R$ while keeping $\mu_F=m$. 
It is clear from the top plot that varying $\mu_F$ and $\mu_R$ independenty 
over the range $m/2$ and $2m$ does not give a wider range of cross section 
values than varying the common scale $\mu=\mu_F=\mu_R$. In fact as can be seen 
from the figure this holds true for a very wide range of scale variation. 
We also note that setting $\mu_F=m/2$ and $\mu_R=2m$ or setting 
$\mu_R=m/2$ and $\mu_F=2m$ still gives a smaller variation than varying 
the common scale $\mu=\mu_F=\mu_R$ between $m/2$ and $2m$. Therefore the NLO 
theoretical uncertainty that we provided above from scale variation 
is not increased by separately varying $\mu_F$ and $\mu_R$. For the approximate
NNLO cross section in the bottom plot of Fig. \ref{toptevmufr} we see that 
the variation with $\mu=\mu_F$ and $\mu_R=m$ affects the upper uncertainty 
(which was stated before as $+0.00$) and this new upper uncertainty 
is $+0.20$. 
However the lower uncertainty ($-0.24$) is unaffected. So the result 
for the approximate NNLO cross section for $m=173$ GeV 
with the scale uncertainy 
from independent $\mu_F$ and $\mu_R$ variation can be written 
as $7.08 {}^{+0.20}_{-0.24} {}^{+0.36}_{-0.27}$ pb. 
Finally, we note that not only is the scale variation with $\mu=\mu_F=\mu_R$ 
greatly reduced in going from NLO to approximate NNLO but so is the separate
$\mu_F$ variation and the separate $\mu_R$ variation.

The MSTW2008 parton densities are the only ones available at NNLO and so we use 
them for our best predictions. It is interesting nevertheless to see if the 
results change significantly using the new CT10 pdf \cite{CT10}, which are at NLO, 
and the pdf errors associated with them. Using CT10 pdf we find a NLO cross section for $m=173$ GeV 
of $6.81 {}^{+0.35}_{-0.75} {}^{+0.42}_{-0.30}$ pb, and an approximate NNLO 
cross section of $7.38 {}^{+0.14}_{-0.25} {}^{+0.45}_{-0.32}$ pb, where the first uncertainty 
is from scale variation (with $\mu_F$ and $\mu_R$ independently varied) and the second 
is from the pdf errors.
We thus find both a larger cross section and a larger uncertainty with CT10 
pdf than with MSTW2008 NNLO pdf.

\subsection{Top quark $p_T$ distribution at the Tevatron}

\begin{figure}
\begin{center}
\includegraphics[width=11cm]{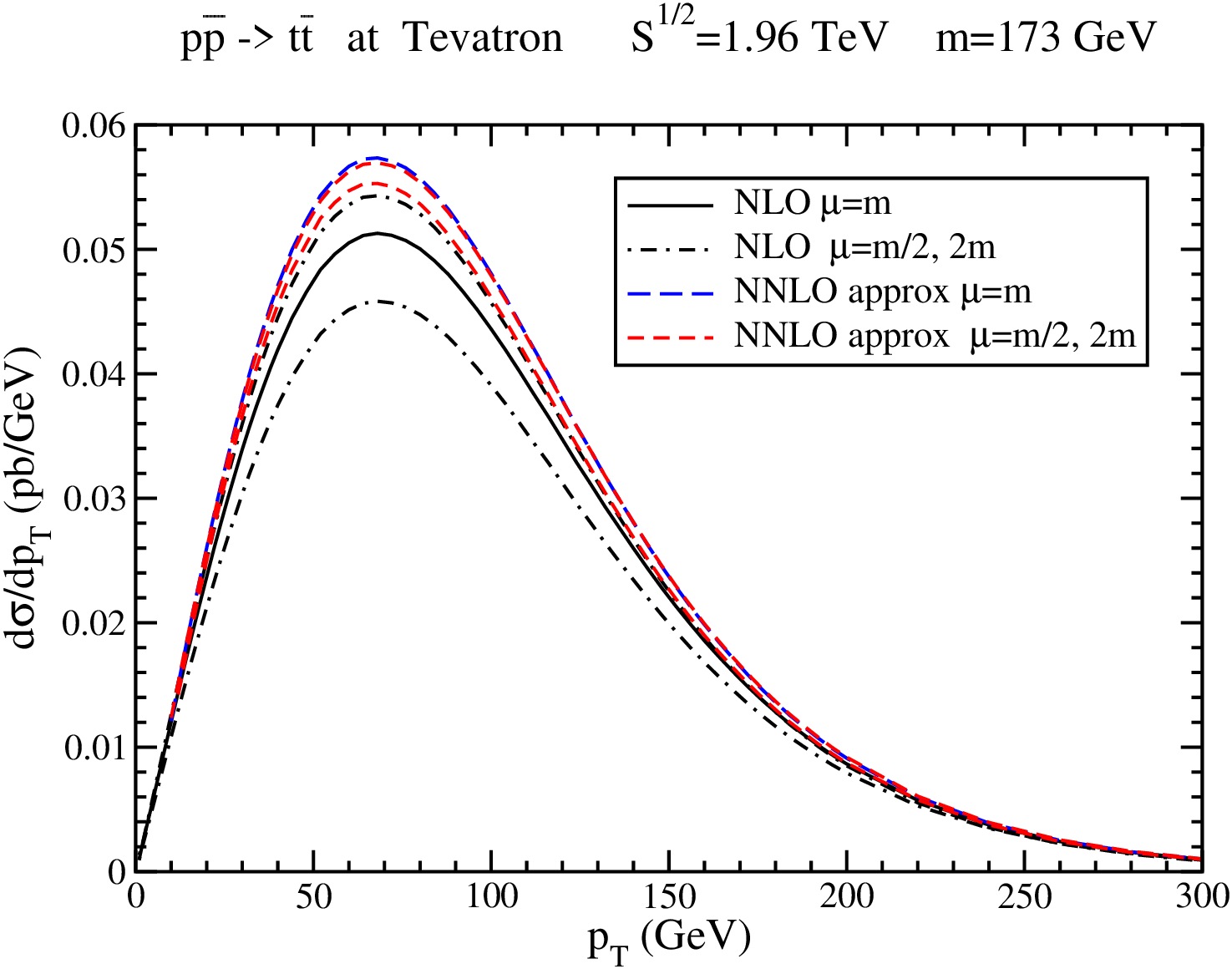}
\caption{The top quark $p_T$ distribution  
at the Tevatron with $\sqrt{S}=1.96$ TeV and $m=173$ GeV.}
\label{pttev1}
\end{center}
\end{figure}

\begin{figure}
\begin{center}
\includegraphics[width=11cm]{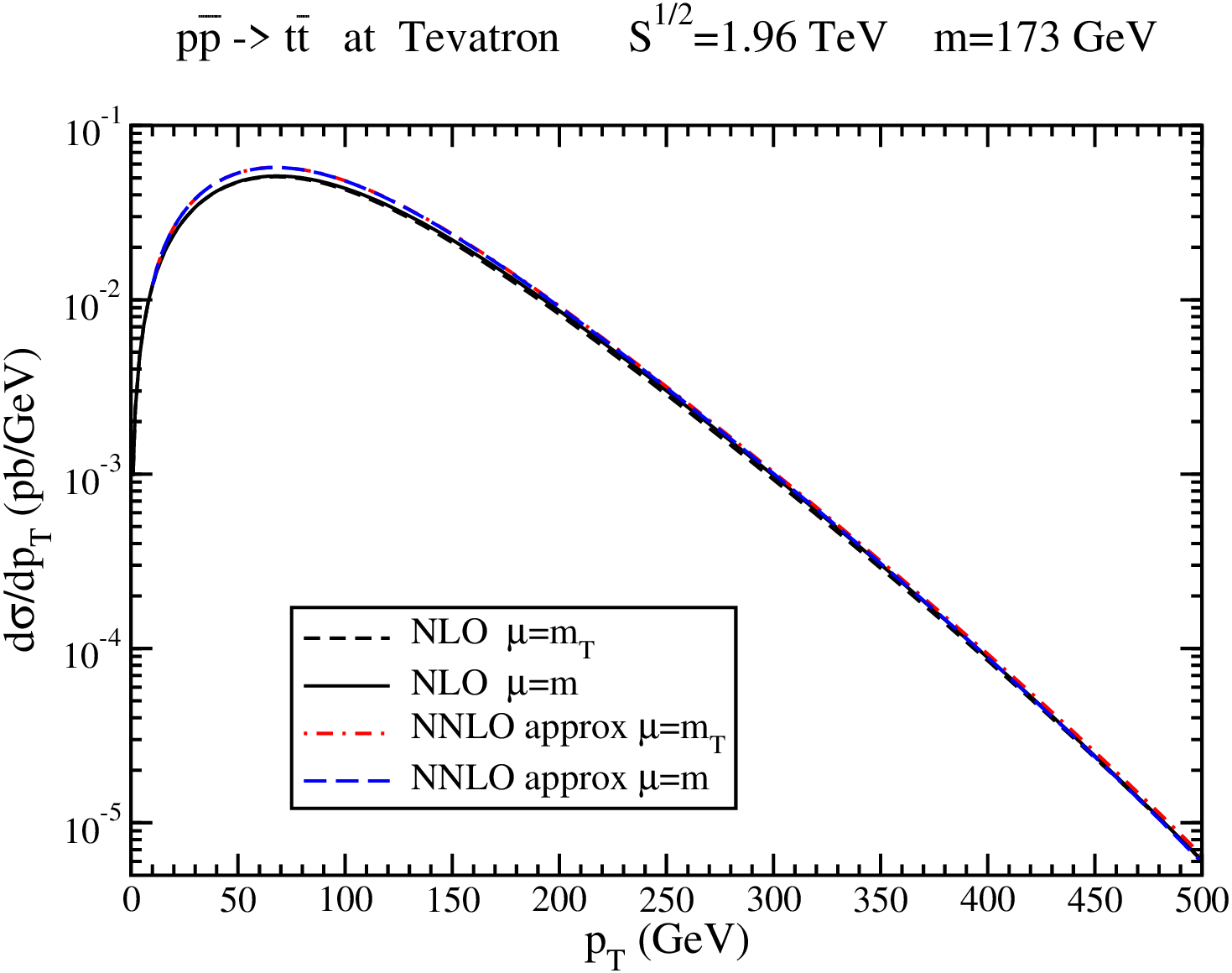}
\caption{The top quark $p_T$ distribution  
at the Tevatron with $\sqrt{S}=1.96$ TeV, $m=173$ GeV, and 
$\mu=m$ or $\mu=m_T$  in a logarithmic plot.}
\label{pttevmTm}
\end{center}
\end{figure}

The top quark transverse momentum distribution at the Tevatron 
with $m=173$ GeV is plotted in Figs. \ref{pttev1} and \ref{pttevmTm}
using the MSTW2008 NNLO pdf. 
Fig. \ref{pttev1} shows the differential distribution $d\sigma/dp_T$ 
over a range $0 \le p_T \le 300$ GeV. Both NLO and NNLO approximate results 
are shown for three different scale choices, $\mu=m/2$, $m$, and $2m$. 
The integrated $p_T$ distribution gives the same result for the total cross 
section as found in the previous subsection, which provides a good consistency 
check of the calculation. The scale variation of the 
$p_T$ distribution at NNLO is again significantly smaller than at NLO.  
The NNLO soft-gluon corrections enhance the NLO result but the shape is 
similar. 

Figure \ref{pttevmTm} presents the top quark $p_T$ distribution 
in a logarithmic plot 
that makes it easier to see $d\sigma/dp_T$ at high $p_T$ values. Results
are now shown up to a $p_T$ of 500 GeV. 
In Fig. \ref{pttev1} the central value for the 
scale was taken to be $\mu=m$ as for the total cross section, and 
the scale variation was around that central value. Another possible 
scale choice for the top quark $p_T$ distribution is the transverse 
mass $m_T$, defined by $m_T=(p_T^2+m^2)^{1/2}$. 
In Fig. \ref{pttevmTm} we show our NLO and approximate NNLO 
results for both $\mu=m$ and $\mu=m_T$.  
We find that the choice of scale, $m$ versus $m_T$, 
makes very little difference even for high $p_T$ of 500 GeV - the curves 
are practically indistinguishable.

Joint threshold and recoil resummation for the $p_T$ distribution (at NLL accuracy only) has 
been studied in \cite{BL}. The effect of recoil is entirely negligible except at 
extremely high $p_T$ ($\sim$800 GeV and above) so we do not consider it further.

\mysection{Top cross section and $p_T$ distribution at the LHC}

We continue with a detailed phenomenological study of top quark 
production in proton-proton collisions at the LHC. We present results for the current LHC energy of 7 TeV 
and the future (design) energy of 14 TeV, and also a few results at 10 TeV.

\subsection{$t{\bar t}$ cross section at the LHC}

\begin{figure}
\begin{center}
\includegraphics[width=11cm]{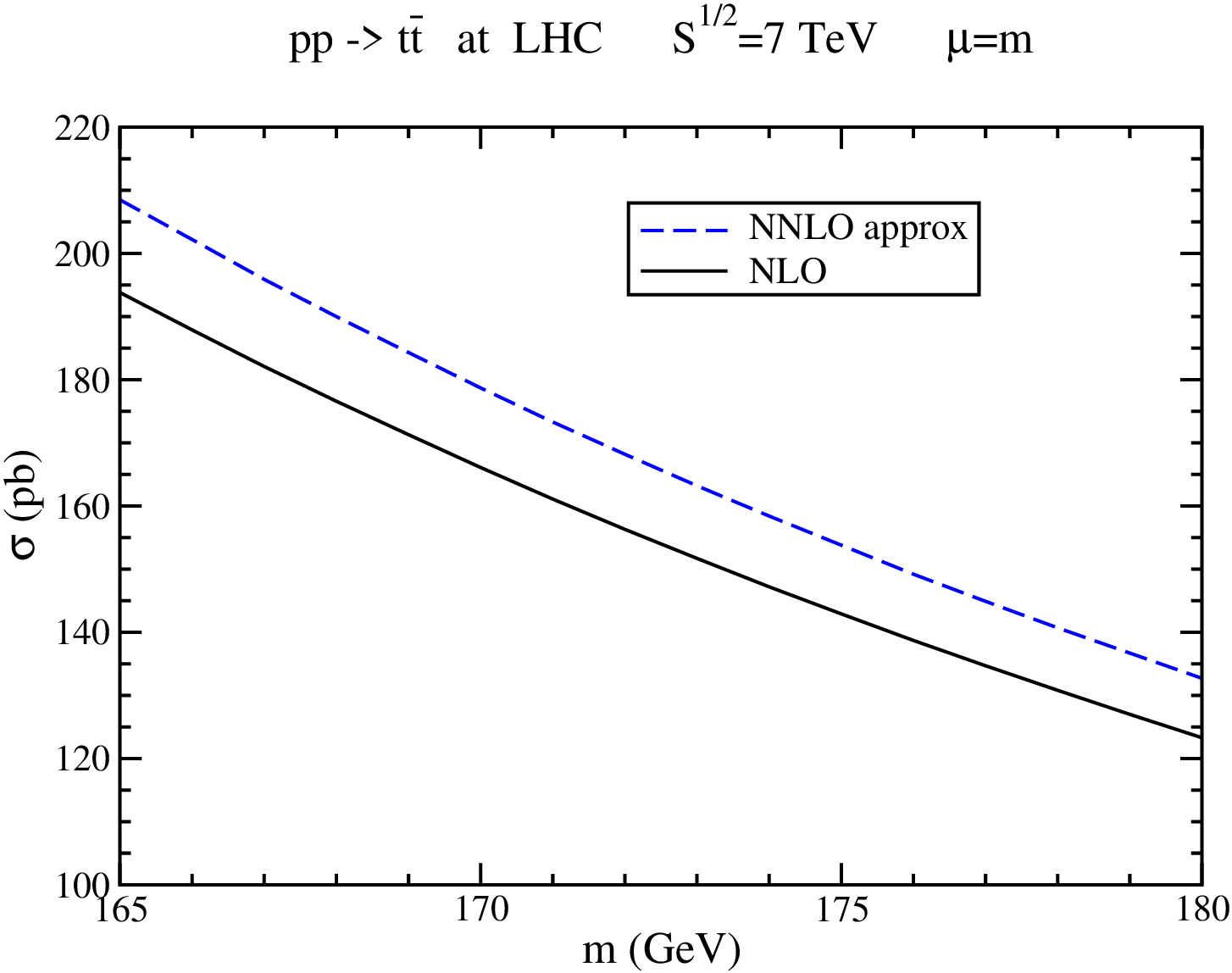}
\caption{The cross section for $t{\bar t}$ production 
at the LHC with $\sqrt{S}=7$ TeV and MSTW2008 NNLO pdf.}
\label{top7lhc}
\end{center}
\end{figure}

In Fig. \ref{top7lhc} we plot the NLO and approximate NNLO cross section 
for top-antitop production at the LHC at 7 TeV energy over a top quark mass 
range $165 \le m \le 180 $ GeV at a factorization and renormalization scale 
$\mu=m$ using the MSTW2008 NNLO pdf. The enhancement from the NNLO soft-gluon corrections is 7.6\%.
Table 1 lists the values for the NNLO approximate cross section at the LHC 
at an enery of 7 TeV for top quark masses between 170 GeV and 175 GeV. 

\begin{figure}
\begin{center}
\includegraphics[width=11cm]{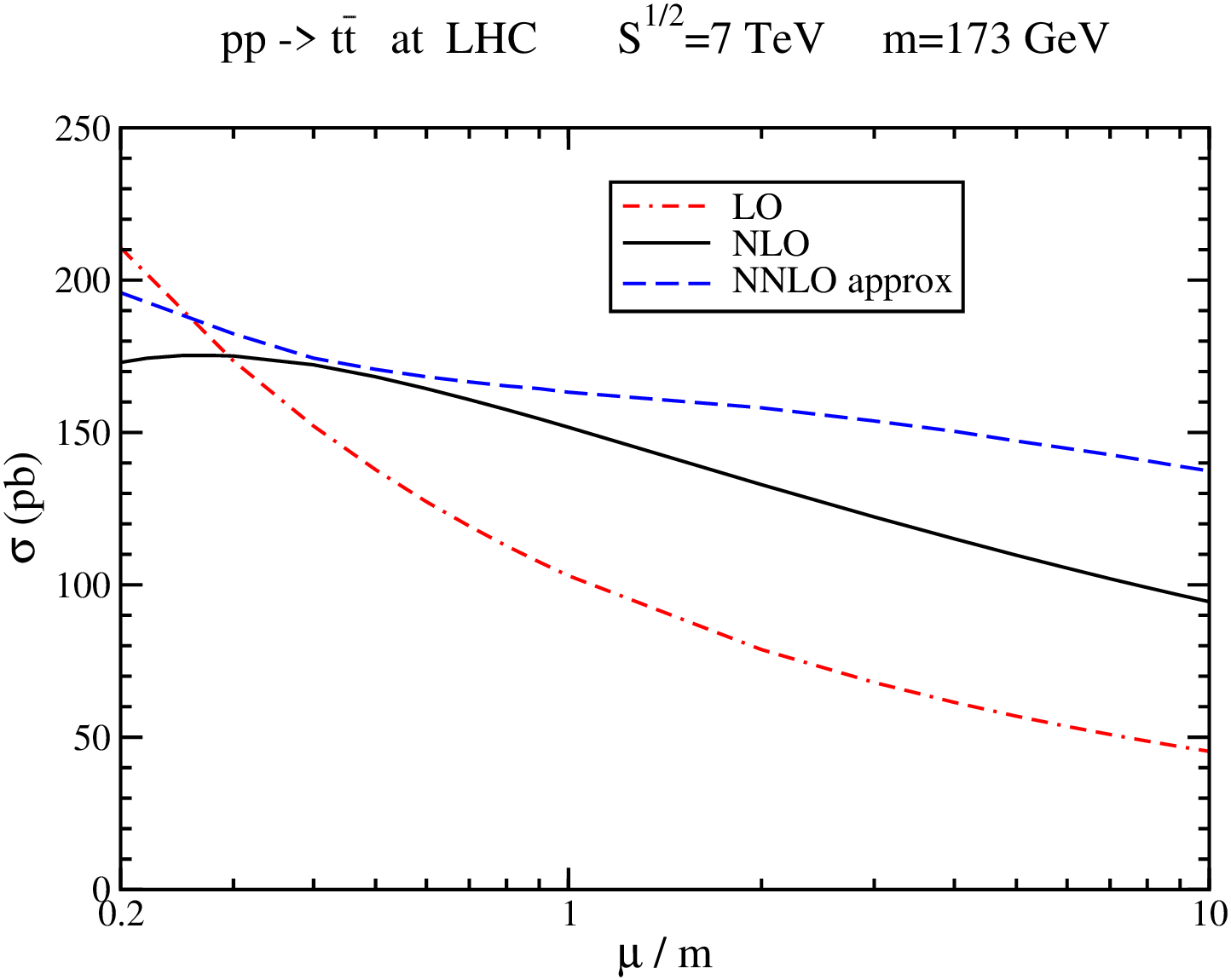}
\caption{The scale dependence of the $t{\bar t}$ cross section  
at the LHC with $\sqrt{S}=7$ TeV and $m=173$ GeV.}
\label{top7lhcmu}
\end{center}
\end{figure}

The scale dependence of the cross section for $m=173$ GeV 
is plotted in Fig. \ref{top7lhcmu}
over a range of two orders of magnitude, $0.2 \le \mu/m \le 10$.  
Again, at LO the cross section is strongly dependent 
on the choice of scale, varying by a factor of 4.64 between maximum and minimum 
values in the range shown. 
The NLO corrections stabilize the LO variation: the NLO cross 
section varies by a factor of 1.85. The NNLO soft-gluon corrections further 
reduce the scale dependence: the NNLO approximate cross section varies 
by a factor of 1.43. The improvement from the NNLO corrections is 
again more impressive if one considers only the traditional variation 
$0.5 \le \mu/m \le 2$. For this range the LO 
cross section varies by a factor of 1.75, the NLO cross section by 1.27, 
while the NNLO approximate cross section by a factor of only 1.08.

For a top quark mass of 173 GeV, the NLO cross section is 
$152 {}^{+16}_{-19}  {}^{+8}_{-9}$ pb 
and the NNLO approximate cross section is 
\beq
\sigma^{\rm NNLOapprox}_{t{\bar t}}(m=173\, {\rm GeV}, \, 7\, {\rm TeV})
=163 {}^{+7}_{-5}  {}^{+9}_{-9} \; {\rm pb}
\eeq
where the first uncertainty is from scale variation over $0.5 \le \mu/m \le 2$ 
and the second is from the MSTW2008 NNLO pdf errors at 90\% C.L.
At NLO the scale uncertainty is about twice as big as the pdf one, but at
NNLO it is significantly smaller.
The scale uncertainty at NNLO is about three times smaller than that at NLO.
Adding the scale and pdf errors in quadrature 
the NNLO approximate result is $163^{+11}_{-10}$ pb, i.e. we have a +7.0\% -6.3\% 
total uncertainty, which is to be contrasted with a much larger (+11.8\% -13.8\%) 
total error at NLO. 

\begin{figure}
\begin{center}
\includegraphics[width=11cm]{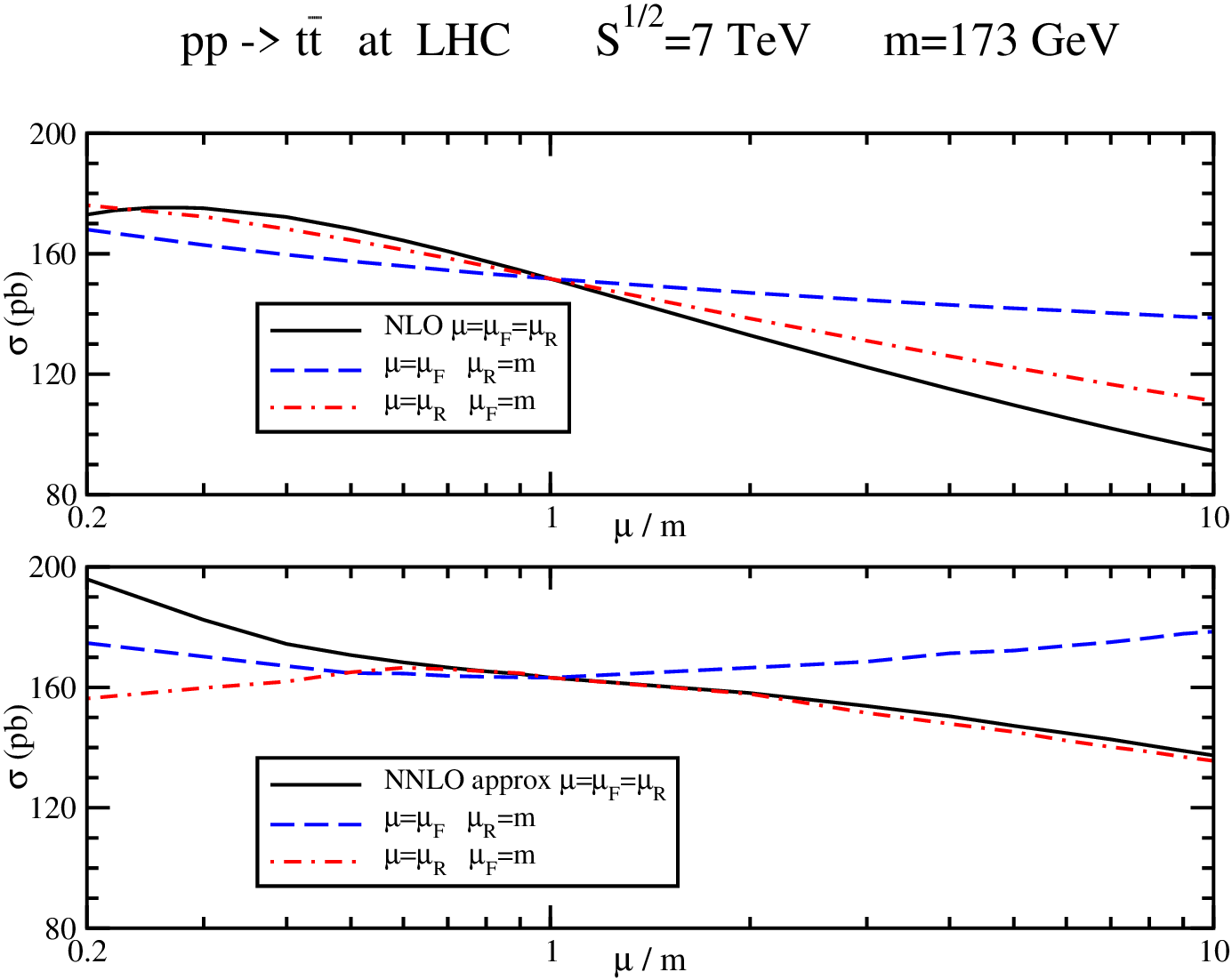}
\caption{The $\mu_F$ and $\mu_R$ dependence of the $t{\bar t}$ cross section  
at the LHC with $\sqrt{S}=7$ TeV and $m=173$ GeV. The top plot 
is at NLO and the bottom is at approximate NNLO accuracy.}
\label{top7lhcmufr}
\end{center}
\end{figure}

We also study the dependence of the cross section separately on the 
factorization scale and the renormalization scale. In Fig. \ref{top7lhcmufr}
we plot the scale dependence of the cross section in three different 
ways at NLO (top plot) and approximate NNLO (bottom plot). The first way 
is to set $\mu=\mu_F=\mu_R$ and vary this common scale, as we did 
in Fig. \ref{top7lhcmu}. The second way is to vary  
$\mu_F$ while keeping $\mu_R=m$, and  
the third way  is to vary $\mu_R$ while keeping $\mu_F=m$. 
From the top plot we see that varying $\mu_F$ and $\mu_R$ independenty 
over the range $m/2$ and $2m$ does not give a wider range of cross section 
values than varying the common scale $\mu=\mu_F=\mu_R$, 
and this actually holds true for nearly the entire wide range of 
scale variation shown in the plot. 
We also find that setting $\mu_F=m/2$ and $\mu_R=2m$ or setting 
$\mu_R=m/2$ and $\mu_F=2m$ still gives a smaller variation than varying 
the common scale $\mu=\mu_F=\mu_R$ between $m/2$ and $2m$. Therefore the NLO 
theoretical uncertainty from scale variation provided previously 
is not increased by separately varying $\mu_F$ and $\mu_R$. For the approximate
NNLO cross section in the bottom plot of Fig. \ref{top7lhcmufr} we also see 
that the independent variation of $\mu_F$ and $\mu_R$ does not affect 
the uncertainty that we wrote previously. 
Finally, we note that the separate $\mu_F$ variation and $\mu_R$ variation are 
reduced when going from NLO to approximate NNLO.

Again we can check if the results change significantly using the 
CT10 pdf, which are at NLO. 
Using CT10 pdf we find a NLO cross section at $m=173$ GeV 
of $150 {}^{+18}_{-20} {}^{+11}_{-10}$ pb, and an approximate NNLO 
cross section of $162 {}^{+9}_{-7} {}^{+12}_{-11}$ pb, so the results 
are very similar to those with MSTW2008 NNLO pdf. 

\begin{figure}
\begin{center}
\includegraphics[width=11cm]{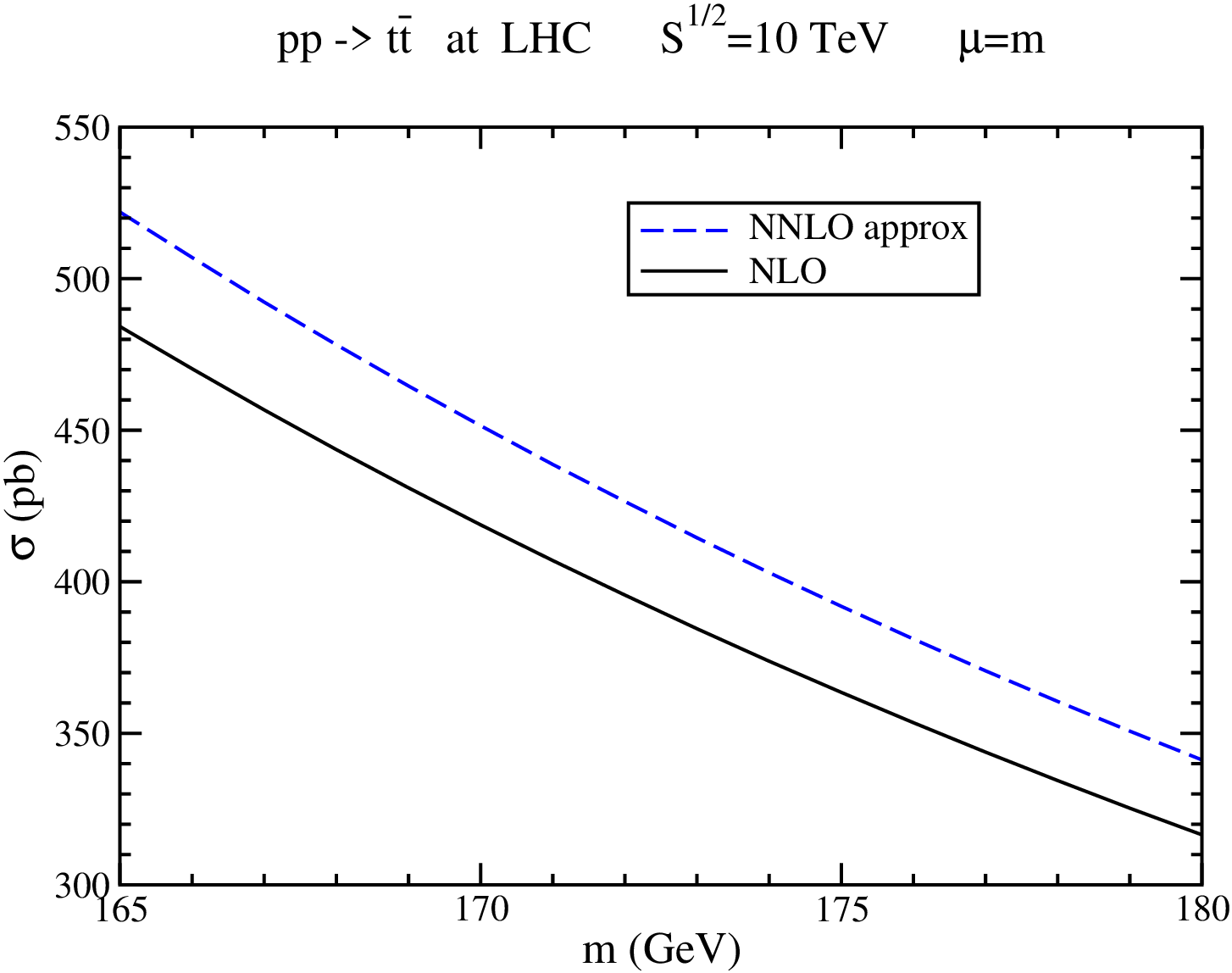}
\caption{The cross section for $t{\bar t}$ production 
at the LHC with $\sqrt{S}=10$ TeV and MSTW2008 NNLO pdf.}
\label{top10lhc}
\end{center}
\end{figure}

For reference, the cross section at a possible future LHC energy of 
10 TeV is plotted 
in Fig.~\ref{top10lhc} using the MSTW2008 NNLO pdf.
For a top quark mass of 173 GeV, we find a NLO cross section of    
$385 {}^{+41}_{-45}  {}^{+17}_{-18}$ pb,
while at NNLO
\beq
\sigma^{\rm NNLOapprox}_{t{\bar t}}(m=173\, {\rm GeV}, 10\, {\rm TeV})
=415 {}^{+17}_{-21}  {}^{+18}_{-19} \; {\rm pb} \, .
\eeq

\begin{figure}
\begin{center}
\includegraphics[width=11cm]{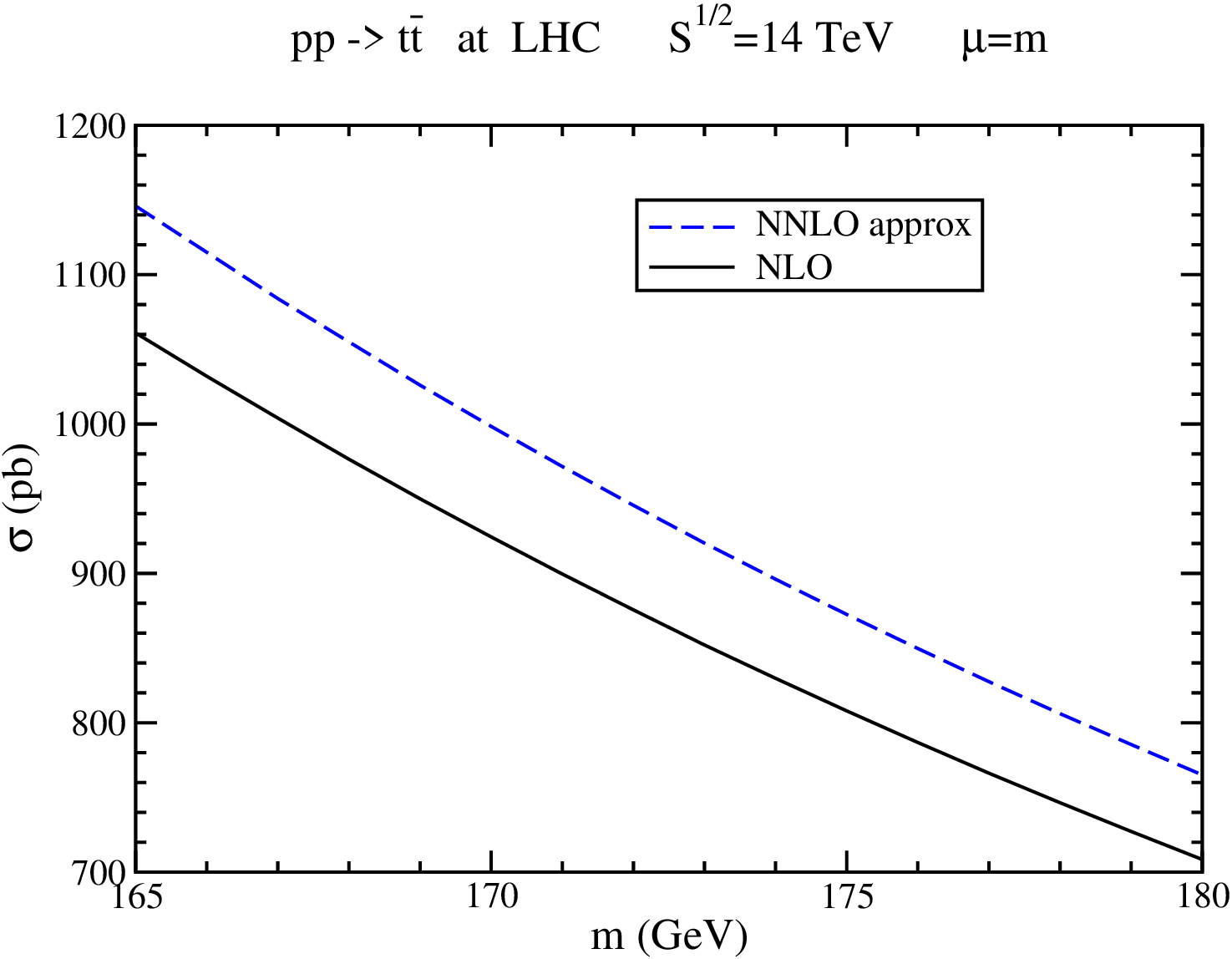}
\caption{The cross section for $t{\bar t}$ production 
at the LHC with $\sqrt{S}=14$ TeV and MSTW2008 NNLO pdf.}
\label{top14lhc}
\end{center}
\end{figure}

The cross section for the design LHC energy of 14 TeV is plotted 
in Fig. \ref{top14lhc} using the MSTW2008 NNLO pdf. The enhancement from the NNLO soft-gluon corrections
is 8.0\%. Table 1 lists the values for the NNLO approximate cross section at 14 TeV 
LHC energy for top quark masses between 170 GeV and 175 GeV. 
The NLO cross section for a top quark mass of 173 GeV is 
$852 {}^{+91}_{-93}{}^{+30}_{-33}$ pb 
and the approximate NNLO cross section is 
\beq
\sigma^{\rm NNLOapprox}_{t{\bar t}}(m=173\, {\rm GeV}, 14\, {\rm TeV})
=920 {}^{+50}_{-39}{}^{+33}_{-35} \; {\rm pb} \, .
\eeq
Again we observe a significant decrease in scale dependence at NNLO relative 
to NLO, and also note that separate variation of $\mu_F$ and $\mu_R$ does not 
increase the uncertainty. The pdf uncertainties at this high energy are much 
smaller than the scale variation at NLO, and somewhat relatively smaller 
at NNLO.
Adding the scale and pdf errors in quadrature 
the NNLO approximate result is $920^{+60}_{-52}$ pb, i.e. we have a +6.5\% -5.7\% 
total uncertainty, which is to be contrasted with a much larger (+11.2\% -11.6\%) 
total error at NLO. 

\subsection{Top quark $p_T$ distribution at the LHC}

\begin{figure}
\begin{center}
\includegraphics[width=11cm]{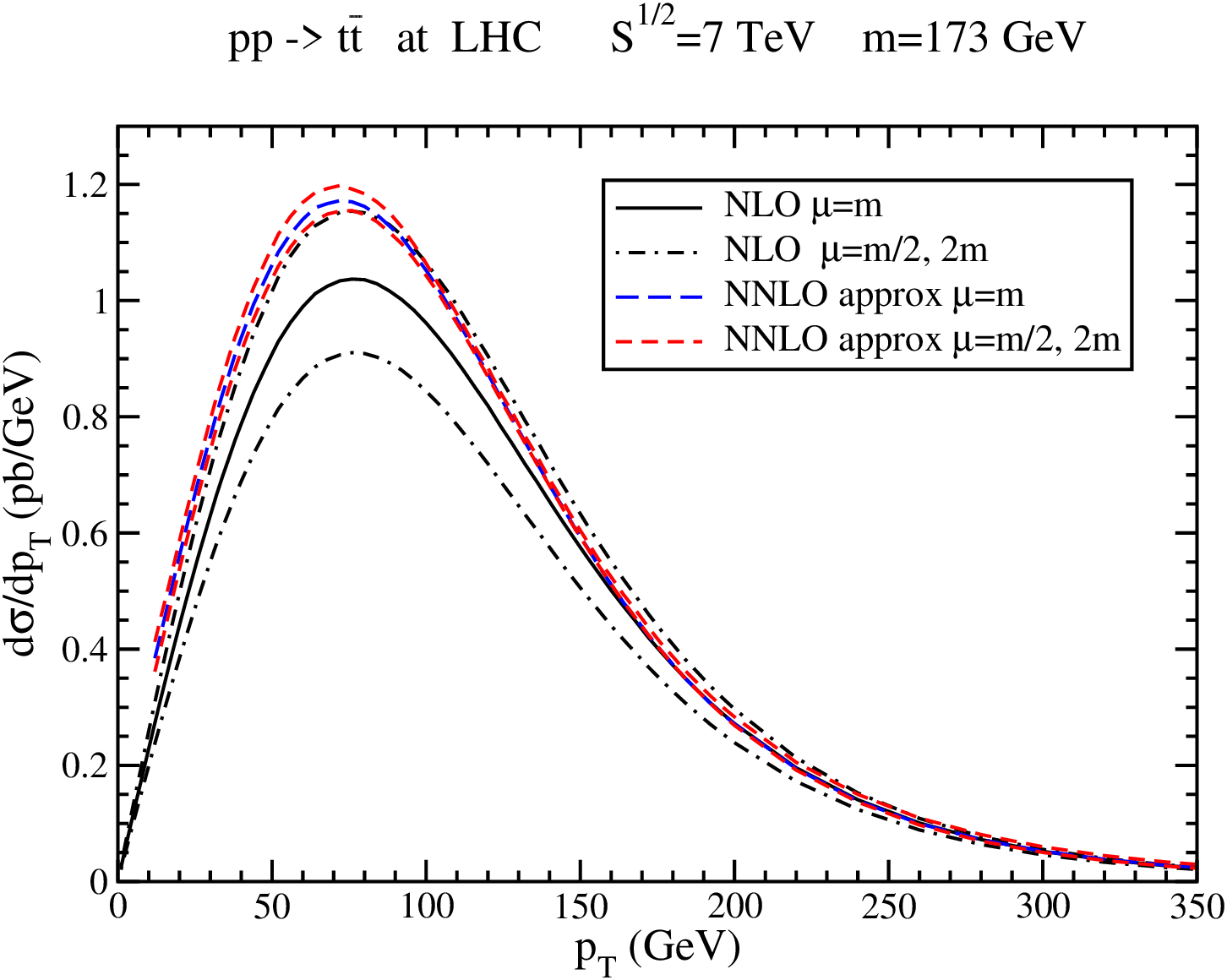}
\caption{The top quark $p_T$ distribution  
at the LHC with $\sqrt{S}=7$ TeV and $m=173$ GeV.}
\label{pt7lhc1}
\end{center}
\end{figure}

\begin{figure}
\begin{center}
\includegraphics[width=11cm]{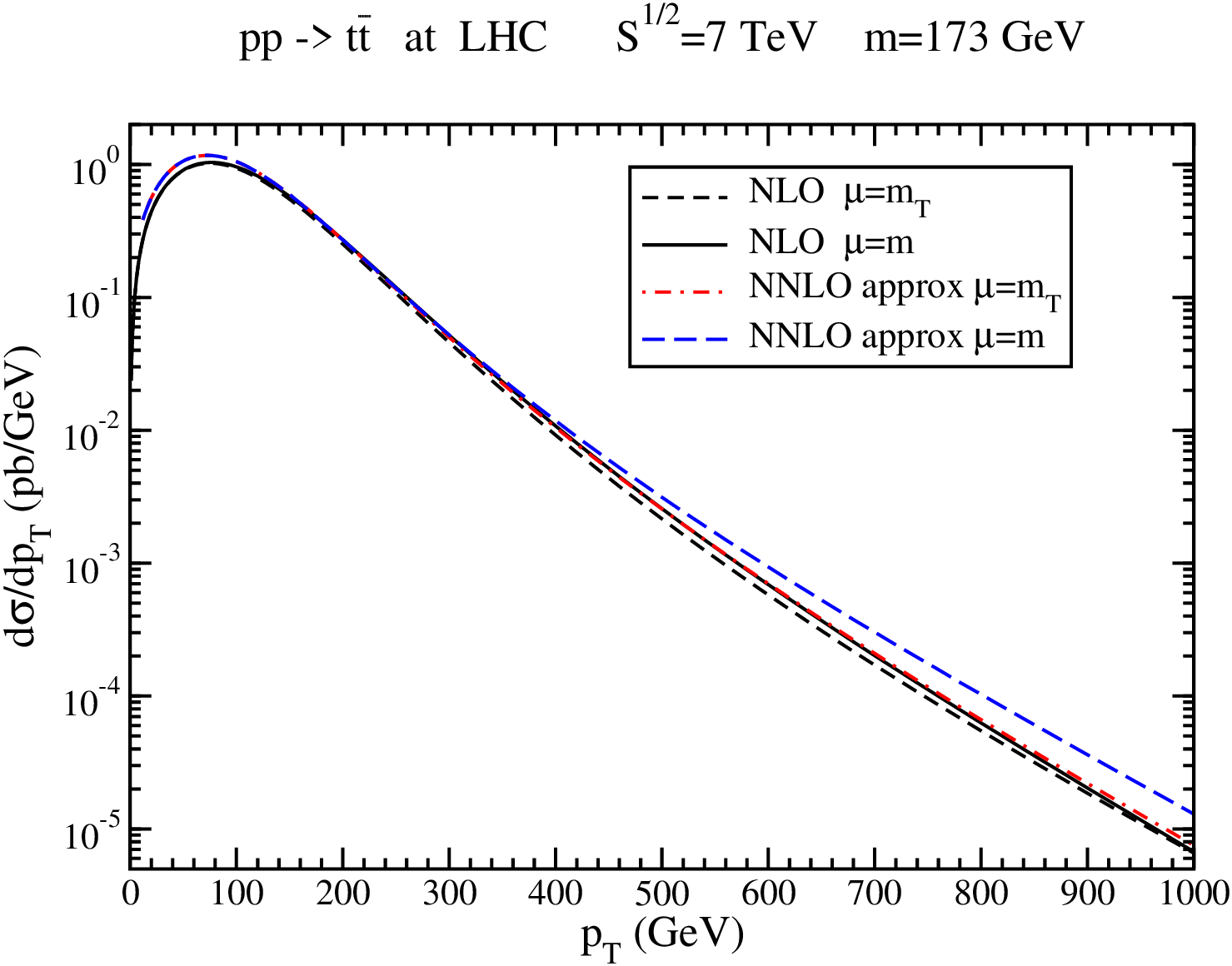}
\caption{The top quark $p_T$ distribution  
at the LHC with $\sqrt{S}=7$ TeV, $m=173$ GeV, and $\mu=m$ or $\mu=m_T$ 
in a logarithmic plot.}
\label{pt7lhcmTm}
\end{center}
\end{figure}

The transverse momentum distribution of the top quark with $m=173$ GeV 
at the LHC at 7 TeV energy is plotted 
in Figs. \ref{pt7lhc1} and \ref{pt7lhcmTm} using the MSTW2008 NNLO pdf. 
Fig. \ref{pt7lhc1} shows NLO and approximate NNLO results for 
the differential distribution $d\sigma/dp_T$ 
over a range $0 \le p_T \le 350$ GeV for three different scale choices, 
$\mu=m/2$, $m$, and $2m$. 
The scale variation of the $p_T$ distribution at NNLO is much  
smaller than that at NLO.  

Figure \ref{pt7lhcmTm} presents the results 
for  $d\sigma/dp_T$ in a logarithmic plot for high $p_T$ values 
up to 1000 GeV, using both $\mu=m$ and $\mu=m_T$, where again $m_T$ is the 
transverse mass. At very high $p_T$ the NNLO approximate corrections become 
increasingly more significant and begin to change the shape of the 
distribution relative to NLO. This is not unexpected since the soft-gluon 
corrections are large near partonic threshold, which is dominant  
at high $p_T$. The change of shape is more 
pronounced with the choice $\mu=m$ than it is with $\mu=m_T$.

\begin{figure}
\begin{center}
\includegraphics[width=11cm]{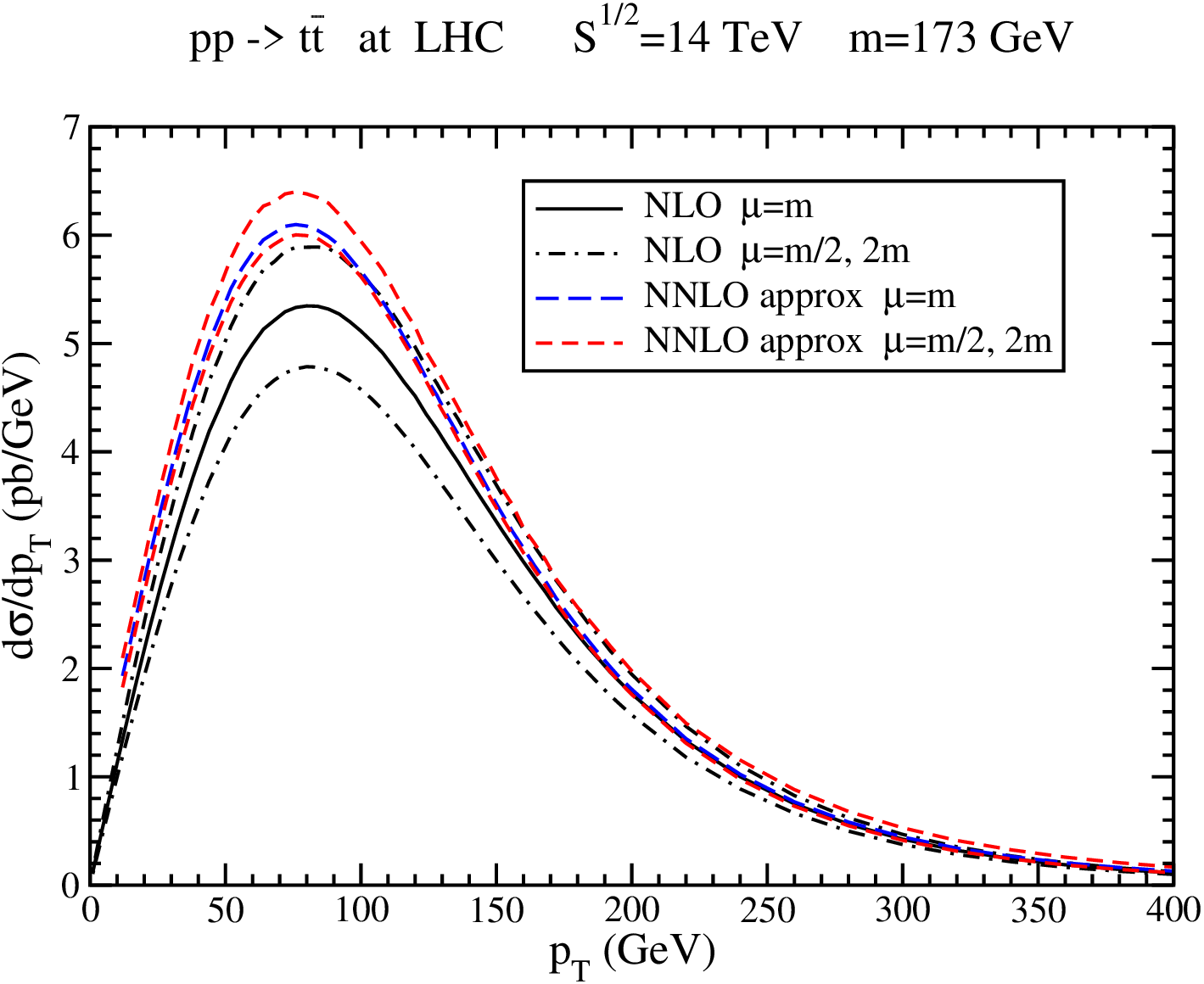}
\caption{The top quark $p_T$ distribution  
at the LHC with $\sqrt{S}=14$ TeV and $m=173$ GeV.}
\label{pt14lhc1}
\end{center}
\end{figure}

\begin{figure}
\begin{center}
\includegraphics[width=11cm]{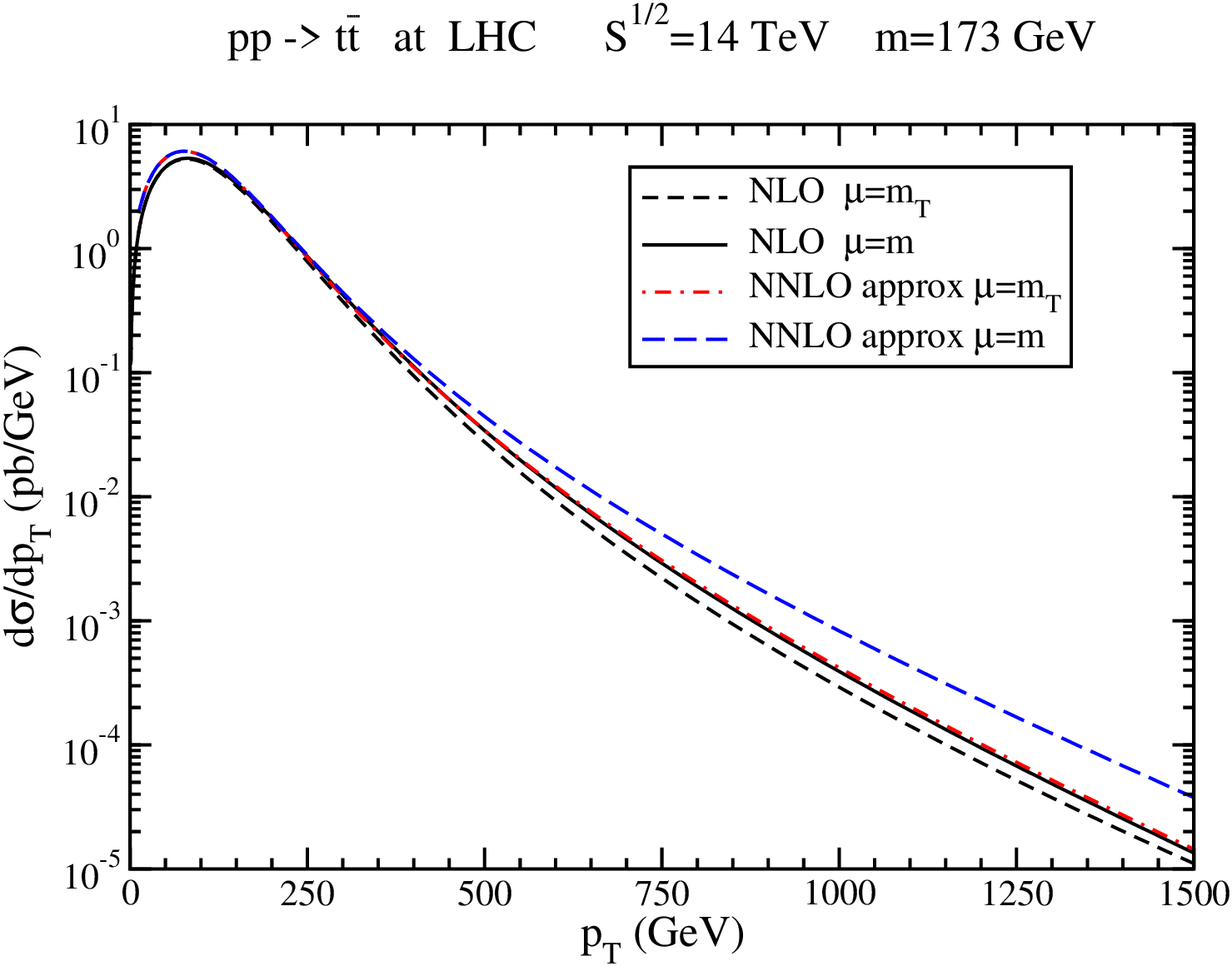}
\caption{The top quark $p_T$ distribution  
at the LHC with $\sqrt{S}=14$ TeV, $m=173$ GeV, and $\mu=m$ or $\mu=m_T$ 
in a logarithmic plot.}
\label{pt14lhcmTm}
\end{center}
\end{figure}

The $p_T$ distribution of the top quark with $m=173$ GeV  
at the LHC at 14 TeV energy is plotted 
in Figs. \ref{pt14lhc1} and \ref{pt14lhcmTm}. 
Fig. \ref{pt14lhc1} shows NLO and approximate NNLO results 
over a range $0 \le p_T \le 400$ GeV for three different scale choices, 
$\mu=m/2$, $m$, and $2m$. 
Again, the scale variation of the $p_T$ distribution at NNLO is much  
smaller than that at NLO.  

Figure \ref{pt14lhcmTm} presents the results for  $d\sigma/dp_T$ 
in a logarithmic plot for high $p_T$ values 
up to 1500 GeV, using $\mu=m$ and $\mu=m_T$.  
The NNLO soft-gluon corrections provide a 
significant enhancement and change the shape of the NLO distribution at 
very high $p_T$. Again, the change in shape is larger with the choice 
$\mu=m$ than it is with $\mu=m_T$.

\mysection{Comparison with other approaches and Conclusions}

In this paper we have resummed the soft-gluon logarithms 
in top quark production to NNLL accuracy. This work directly 
extends the earlier NLL resummation in Ref.~\cite{NKGS} and the further  
work in \cite{NK01,KLMV} and later in \cite{NKRV1,NKRV2}. 
To achieve NNLL (NLL) accuracy, it is necessary to derive the soft anomalous 
dimension matrix at two (one) loops for each partonic process. 
At NLL and NNLL the color structure of the hard scattering enters the 
resummation in a non-trivial way.
The soft anomalous dimension matrices are explicitly dependent on the 
kinematical variables $s$, $t_1$, $u_1$, and the resummation involves 
these quantities and logarithms of $s_4$, where $s_4=s+t_1+u_1$ measures distance from partonic threshold.
Thus this is a fully differential calculation and the formalism 
in this paper has been used to calculate 
not only total cross sections but also differential cross sections, 
such as transverse momentum distributions. 
Approximate NNLO differential cross sections are extracted from the 
resummation (higher-order contributions beyond NNLO are small, 
see e.g. Ref. \cite{NNNLO}). The NNLO expansion avoids the need 
for prescriptions to deal with Landau-pole divergences in the resummation, 
and we prefer to take this approach
since the numerical discrepancies between different prescriptions
are larger than the corrections beyond NNLO  
(see e.g. discussion in \cite{NK01,NKRV2}).

There also exist formalisms of resummation and finite-order expansions for 
the total cross section only \cite{BCMN} that are calculationally simpler, 
and only involve the variable $\beta=\sqrt{1-4m^2/s}$. Logarithms of $\beta$ 
have been resummed at NLL in 
\cite{BCMN,CFMNR} and at (partial) NNLL in \cite{MU} 
(Ref. \cite{MU} made an incorrect assumption about the 
two-loop terms which, as later understood, is not valid). This approach does  
not, however, involve the exact differential kinematics and hence numerical 
deviations from the exact kinematics-sensitive result may appear. 
Furthermore this approach is inapplicable to $p_T$ or other differential 
distributions, so it is limited in scope. For further discussion 
of the differences see also Ref. \cite{NKRV2} and \cite{AFNPY}. 
More recently, complete NNLL results in logarithms of $\beta$ have appeared in 
Ref. \cite{BFS} using soft-collinear and non-relativistic effective theory, 
and in Ref. \cite{CMS} using resummation 
in moment space.  Threshold expansions to NNLO for the total cross section 
from this $\ln\beta$ resummation have recently appeared in \cite{BCFMS,BFKS}. 
Again, all these results are for total cross sections only, based on expansions in 
$\beta$. It is important to note that the terminology ``NLL'' and ``NNLL'' means different things 
in the approaches of Refs. \cite{BFS,CMS,BCMN,CFMNR,MU,BCFMS,BFKS} than it does in  
the differential-level formalism of Refs. \cite{NKRV1,NKRV2,NKGS,NK01,KLMV} and of this paper because different types of logarithms are resummed.  

Another differential-level  formalism that has recently 
appeared \cite{AFNPY,AFNPYex} is based on 
soft-collinear effective theory and heavy-quark effective theory. 
While the resummation in \cite{NKRV1,NKRV2,NKGS,NK01,KLMV} and this paper is done in moment space, in 
\cite{AFNPY} it is performed in momentum space. The total cross section 
and invariant mass distribution at NNLL have been presented in 
\cite{AFNPY}. The total cross section results in \cite{AFNPY} are quite 
different from those in this paper.  
One major reason for the difference is the different choice of kinematics, 
as we describe below.

In Refs. \cite{NK01,KLMV,NKRV1,NKRV2} (based on the formalism of 
\cite{NKGS}) results were provided in both 
single-particle-inclusive (1PI) 
and pair-invariant-mass (PIM) kinematics.  The kinematics 
ambiguity was studied in detail in \cite{KLMV} and found to be
an important source of uncertainty. 
In 1PI kinematics the soft-gluon logarithms are of the form 
$[\ln^k(s_4/m^2)/s_4]_+$ and the soft-gluon corrections to the 
double differential cross section, $d^2\sigma/(dt_1 du_1)$, are calculated. 
In PIM kinematics, the soft logarithms are of the form $[\ln^k(1-z)/(1-z)]_+$ 
with $z=M^2/s$, and $z\rightarrow 1$ near threshold, 
where $M^2$ is the $t \overline t$ pair mass squared. 
In PIM kinematics, the soft gluon corrections to the 
double differential cross section, $d^2\sigma/(dM^2 d\cos\theta)$, where 
$\theta$ is the scattering angle in the partonic center-of-mass frame, 
are calculated.  The cross section in PIM kinematics was found to be smaller 
than the 1PI result.  
The results in Refs.~\cite{NK01,KLMV} were based on NLL resummation and 
were later improved by adding subleading terms \cite{NKRV1,NKRV2}.
The kinematics ambiguity was thus reduced in \cite{NKRV1,NKRV2}. Still it 
was shown in \cite{NKRV1} that the PIM kinematics gives large 
negative results at NNLO for the $gg$ channel at LHC energies (for $t{\bar t}$ 
production at the LHC, 
the $gg$ channel is by far dominant over the $q{\bar q}$ channel). 
These negative corrections are deemed unphysical since 
already at NLO the PIM approximation for the corrections does not reproduce 
well the exact NLO result while the 1PI result is a much better approximation 
(detailed 
graphs for the partonic scaling functions in 1PI and PIM kinematics were shown 
in \cite{KLMV} and also \cite{NKRV1}). In the present 
paper we have thus used 1PI kinematics. In contrast, Ref. \cite{AFNPY} uses 
a modified PIM kinematics. Although the modified PIM kinematics of Ref. \cite{AFNPY} 
produces less negative results than PIM in \cite{KLMV} and \cite{NKRV1}, the overall NNLO contribution 
in modified PIM is still negative. 
This explains why both the NNLL resummed cross section and the NNLO approximate cross section with modified PIM in \cite{AFNPY} is less 
than the NLO cross section at $\mu=m$ for both Tevatron and LHC energies. 
This is in sharp contrast to the 1PI results here and in all our previous calculations
(at both NLL and NNLL accuracy) where the NNLO soft-gluon corrections are found to 
provide a positive enhancement of the NLO cross section. The 1PI kinematics 
provides an excellent approximation as evidenced by the fact that the NLO 
approximate 1PI corrections from the expansion of the resummed cross section 
account for well over 98\% (up to 99.8\%) of the exact NLO corrections in the $gg$ channel (with $\mu=m$) at both Tevatron and LHC energies. This is a far better agreement than can be attained with PIM or modified PIM kinematics. 
We thus remain of the opinion that the results in \cite{AFNPY} do not accurately reflect the true contribution of soft-gluon corrections.   
 
It is also interesting to compare the results in this paper with our previous  
results in \cite{NKRV1,NKRV2}.
Although NNLL resummation requires calculation of the two-loop 
soft anomalous dimension matrices as presented in this paper, it was argued 
in \cite{NKRV1,NKRV2} that the numerical contribution of this matrix at 
two loops to the cross section is expected to be small.
In \cite{NKRV1,NKRV2} many of the terms beyond NLL were already included 
in the calculation 
and it was argued based on the study of the scaling functions in 1PI and PIM 
kinematics that these additional subleading terms were relatively dominant. 
Now that the full two-loop NNLL terms are known it is important to revisit the  
validity of this argument. We find that indeed the new two-loop terms from the
soft anomalous dimension matrices contribute very little to the total cross section, 
and hence the argument was valid and the results in \cite{NKRV1,NKRV2} were 
robust. For example, using the MSTW2008 NNLO pdf \cite{MSTW2008} the 
calculation at the accuracy of Ref. \cite{NKRV2} for the top quark cross 
section at the LHC at 7 TeV gives 165 pb, while in this
paper we find 163 pb based on NNLL resummation. The difference 
between the two numbers is very small compared with the overall 
theoretical uncertainty. Any differences in the numbers provided in 
\cite{NKRV1,NKRV2}, and the present work are overwhelmingly due 
to the use of different pdf and only to a rather small extent due to the 
different theoretical accuracy. 

To conclude, we have shown in this paper that the top quark cross section and 
transverse momentum distribution receive  significant enhancements from soft-gluon 
corrections at NNLO. These corrections have been resummed at NNLL accuracy by calculating 
the two-loop soft anomalous dimension matrices for the partonic processes.
Approximate NNLO total and differential cross sections have been derived from 
the NNLL resummed result. The NNLO soft-gluon corrections enhance the total 
cross section and the $p_T$ distribution and greatly reduce the theoretical 
uncertainty from scale variation. The pdf uncertainty of the cross section has 
also been presented. Our NNLL resummation formalism can be used 
to calculate other differential distributions of interest, such as the top  
quark rapidity distribution. This will be a topic of future work.

\mysection*{Acknowledgements}
This work was supported by the National Science Foundation under 
Grant No. PHY 0855421.

\end{document}